\documentclass[reprint,twocolumn,superscriptaddress,showpacs,floatfix]{revtex4-1}

\usepackage{graphicx}
\usepackage{epsfig}
\usepackage{amsmath}
\usepackage{exscale}
\usepackage{array}
\usepackage{units}
\usepackage{amssymb}
\usepackage{setspace}
\usepackage{color}
\usepackage{enumitem}
\definecolor{lightgrey}{rgb}{0.87,0.87,0.87}

\providecommand{\e}[1]{\ensuremath{\times 10^{#1}}}

\begin{document}

\title{Exploring connections between statistical mechanics and Green's functions for realistic systems. Temperature dependent  electronic entropy and internal energy from a self-consistent second-order Green's function.}
\author{Alicia Rae Welden}
\email[Corresponding author: ]{welden@umich.edu}
\affiliation{Department of Chemistry, University of Michigan, Ann Arbor, Michigan 48109, USA}
\author{Alexander A. Rusakov}
\email[Corresponding author: ]{rusakov@umich.edu}
\affiliation{Department of Chemistry, University of Michigan, Ann Arbor, Michigan 48109, USA}
\author{Dominika Zgid}
\email[Corresponding author: ]{zgid@umich.edu}
\affiliation{Department of Chemistry, University of Michigan, Ann Arbor, Michigan 48109, USA}

\begin{abstract}

Including finite-temperature effects from the electronic degrees of freedom in electronic structure calculations of semiconductors and metals is desired; however, in practice it remains exceedingly difficult when using zero-temperature methods, since these methods require an explicit evaluation of multiple excited states in order to account for any finite-temperature effects. Using a Matsubara Green's function formalism remains a viable alternative, since in this formalism it is easier to include thermal effects and to connect  the dynamic quantities such as the self-energy with static thermodynamic quantities such as the Helmholtz energy, entropy, and internal energy.  However, despite the promising properties of this formalism, little is know about the multiple solutions of the non-linear equations present in the self-consistent Matsubara formalism and only a few cases involving a full Coulomb Hamiltonian were investigated in the past. Here, to shed some light onto the iterative nature of the Green's function solutions, we self-consistently evaluate the thermodynamic quantities for a one-dimensional (1D) hydrogen solid at various interatomic separations and temperatures using the self-energy approximated to second-order (GF2). At many points in the phase diagram of this system, multiple phases such as a metal and an insulator exist, and we are able to determine the most stable phase from the analysis of Helmholtz energies. Additionally, we show  the evolution of the spectrum of 1D boron nitride (BN) to demonstrate that GF2 is capable of qualitatively describing the temperature effects influencing the size of the band gap.

\end{abstract}
\maketitle
\section{Introduction}
%
In molecular quantum-chemical calculations of thermodymanic properties such as Gibbs energy~\cite{mcquarrie1999molecular,g09thermo}, the temperature dependent component is usually dominated by vibrational contributions. This is due to the large gaps between electronic states, which ensure that the excited state populations will be negligible. Consequently, common molecular calculations do not explicitly include temperature effects on the electronic structure.


However, for materials such as doped semiconductors~\cite{guidebookworld} the magnitude of the electronic band gap can be relatively small, or for metals nonexistent altogether, allowing electronic states other than the ground state to be accessible at low temperatures. Thus, it is necessary to include temperature effects into the electronic description. 
Even though for most materials the vibrational contribution to the specific heat is much larger than the electronic one, there are cases when the incorporation of the electronic contribution is necessary. The electronic contribution to specific heat is important for {\bf (i)} heavy fermion materials at low temperatures \cite{metalsfreeenergy}, {\bf (ii)}  materials that do not undergo a structural transition that changes the vibrational contribution, causing the stability of phases to primarily depend on the relative electronic contribution to Gibbs energy, {\bf (iii)} materials with a structural transition where the difference between vibrational contributions of the phases is of the same order as the difference between electronic contributions \cite{benzenetuckerman,resorcinol}. Thus, modern materials calculations can benefit from computational tools that provide access to the temperature dependent electronic contribution to the specific heat, Gibbs energy, entropy, or electronic part of the partition function.

 While in traditional quantum chemical calculations evaluating the electronic contribution to temperature dependent quantities is certainly not wide spread, a number of such methods exist, most notably, the 
 finite temperature Hartree-Fock (HF) \cite{mermin1963stability}, density functional theory (DFT) \cite{mermin1965thermal,dharma2016current, smith2015thermal,trickey},  M\o ller-Plesset second order perturbation theory (MP2) \cite{hirata2013kohn}, coupled-cluster (CC) \cite{altenbokum1987temperature,mandal2003finite,sohiratacc},  and Lanczos method \cite{jaklivc1994lanczos, ftlanczos} for finite temperature configuration interaction (CI) calculations.
 However, these methods are usually quite difficult to implement and costly to use since they rely on the modification of the parent zero-temperature method to the finite temperature formalism by adapting it to work in the canonical or grand canonical ensemble. For example, to carry out such calculations in the CI formalism, one must obtain the excited states and corresponding Boltzmann factors in order to evaluate the partition function and thermodynamic averages, making application of finite temperature variants of these methods quite cumbersome.
 
Conversely, for the Green's function formalism the connection to thermodynamics arises in a straightforward manner and was derived in numerous books in the past~\cite{vanleeuwentext,mattuck2012guide,fetter2003quantum,dzyaloshinski1975methods}. While this theoretical connection is well understood, the actual numerical calculations of the thermodynamic quantities still remain quite challenging since a fully self-consistent imaginary axis (Matsubara) Green's function calculation is desired. A non-self-consistent Green's function can result in non-unique thermodynamic quantities.


The  fully self-consistent Green's function calculations are challenging for multiple reasons, such as large imaginary time and frequency grids required for convergence or multiple solutions that can be present due to the non-linear nature of the equations. It is for reasons such as these that in recent years calculations that capitalized on Green's function language and yield thermodynamic quantities where mostly done for model systems~\cite{potthoff1,potthoff2,potthoff2003Hubbard,PotthoffSE,potthoff2003self}. 

Here, we would like to stress that while multiple large scale real axis Green's function calculations are performed at present for the single shot $G_0W_0$, $GW_0$, or semi-self consistent $GW$ for large realistic systems~\cite{neuhauser2014breaking, hung2016excitation, van2006quasiparticle, caruso2013self, faleev2004all, rostgaard2010fully, govoni2015large}, currently, only a few research groups  have managed to rigorously generalize the self-consistent finite temperature (imaginary axis) Green's function formalism to deal with a general Hamiltonian containing all the realistic interactions~\cite{van2006total,dahlen2006variational,GF2,rusakov,spline,AlexeiSEET}. Thus, any insight gained from studying even simple periodic systems and analyzing the possible self-consistent solutions of the Matsubara formalism remains valuable.


To the best of our knowledge, here, we present the first application of the fully self-consistent finite temperature Green's function formalism to evaluate thermodynamic quantities and phase stability for a periodic system described by a full quantum chemical Hamiltonian. We demonstrate that Green's function formalism leads to a simple calculation of the electronic contribution to the Helmholtz or Gibbs energy, entropy, grand potential, and partition function without explicitly performing any excited state calculations. The presented formalism is exact at the infinite temperature limit since the perturbative Green's function formulation is a perturbation that contains the inverse temperature as small parameter. 


 This paper is organized as follows. In Sec.~\ref{sec_thermo_con}, we introduce the imaginary Green's function formalism and its connection to thermodynamics. In Sec.~\ref{sec_phi}  and~\ref{sec_eval_phi_gf2}, we list properties of the Luttinger-Ward functional which is our main computational object and we explain its evaluation within a self-consistent Green's function second-order (GF2) periodic implementation.
In Sec.~\ref{sec_results}, we present numerical results first for a benchmark molecular problem and then for two 1D systems: periodic hydrogen as well as boron nitride. Finally, we form conclusions in Sec.~\ref{sec_conclusions}.

 \section{Connection with thermodynamics}\label{sec_thermo_con}

One of the first descriptions of the connection between the Green's function formalism and thermodynamics was presented in the book by Abrikosov, Gorkov, and Dzyaloshinski~\cite{dzyaloshinski1975methods}. Since then multiple texts have appeared that discuss this connection~\cite{vanleeuwentext,fetter2003quantum,vanneck}. For an excellent, detailed derivation, we encourage the reader to follow Ref.~\onlinecite{vanleeuwentext};
 here, we will only mention few basic Green's function equations for the sake of completeness.  
 
 The one-body Green's function is defined as 
 \begin{eqnarray}\label{GF_general}
 G_{ji}(z_1,z_2)\equiv &
 \frac{{\rm Tr}[e^{-\beta \hat{H}^{M}} \hat{G}_{ji}(z_1,z_2)]}{{\rm Tr}[e^{-\beta \hat{H}^{M}}]}\\
 = &\frac{1}{i} \frac{{\rm Tr} [\tau \{e^{-i\int_{\gamma }d\bar{z}\hat{H}(\bar{z})}\hat{d}_{j,H}(z_1)\hat{d}^{\dagger}_{i,H}(z_2)\}]}
 {{\rm Tr}[\tau \{e^{-i \int_{\gamma } d\bar{z}\hat{H}(\bar{z})} \} ]} \nonumber,
 \end{eqnarray}
 where $\hat{d}_{j,H}(z_1)$ and $\hat{d}^{\dagger}_{i,H}(z_2)$ are the second-quantized annihilation and creation operators in the Heisenberg representations and $\beta=1/(k_B T)$ is the inverse temperature while $T$ is the actual temperature and $k_B$ is the Boltzmann constant.
 This Green's function, depending on how the contour $\gamma$ in the complex plane is closed, can be used to describe system's  time evolution (when $z_1$ and $z_2$ are set to the real-time variables), zero temperature phenomena, 
 or  equilibrium phenomena at finite temperature.
 
 Here, we are interested in a formalism used to calculate the initial ensemble average that is applied to systems in thermodynamic equilibrium at finite temperature. This approach is called Matsubara formalism or the ``finite-temperature formalism".
 For this reason, in the Green's function from Eq.~\ref{GF_general}  we set
 $z_1=t_0-i\tau_1$ and $z_2=t_0-i\tau_2$. Consequently, the Green's function 
 \begin{eqnarray}\label{GF_Matsubara}
 G_{ji}^{M}(\tau_1,\tau_2& ) 
 = 
  \frac{1}{i}
 \big\{ 
 \theta ( \tau_1-\tau_2 )
 \frac{ {\rm Tr} [e^{(\tau_1-\tau_2-\beta)\hat{H}^{M}} \hat{d}_{j}e^{(\tau_2-\tau_1) \hat{H}^{M}}\hat{d}^{\dagger}_{i} ] } { {\rm Tr}[e^{-\beta \hat{H}^{M}}]}   \nonumber  \\
& \pm \theta ( \tau_2-\tau_1 ) 
 \frac{ {\rm Tr} [e^{(\tau_2-\tau_1-\beta)\hat{H}^{M}} \hat{d}^{\dagger}_{i}e^{(\tau_1-\tau_2) \hat{H}^{M}}\hat{d}_{j} ] } { {\rm Tr}[e^{-\beta\hat{H}^{M}}]} 
  \big\}
  \end{eqnarray}
 does not describe any time evolution of the system under study. Instead, in this Green's function the initial state of the system can be the thermodynamic state corresponding to a Hamiltonian, $\hat{H}^{M}=\hat{H}(t_0)-\mu \hat{N}$, where $\hat{N}$ is the particle number operator.
 Thus, from this Matsubara Green's function, the initial ensemble average of any one-body operator $\hat{O}=\sum_{ij} O_{ij} \hat{d}^{\dagger}_{i} \hat{d}_{j}$, can be evaluated simply as
 \begin{eqnarray}
 O=\frac{{\rm Tr} [ e^{-\beta\hat{H}^{M}}\hat{O}]} {{\rm Tr} [e^{-\beta\hat{H}^{M}}]}=\pm i \sum_{ij}O_{ij}G^{M}_{ji}(\tau),
   \end{eqnarray}
where $\tau=\tau_1-\tau_2$ since the one-body Green's function matrix elements depend only on the difference of the imaginary time variables. Furthermore, the imaginary time Green's function $G(\tau)$
can be Fourier transformed to the imaginary frequency Green's function $G(i\omega_n)$ where for fermions the frequency grid is given by $\omega_n=\frac{(2n+1)\pi}{\beta}$ with $n$ defined here as a positive integer. 
For simplicity, in the remainder of this paper,  we will drop subscript $M$ denoting Matsubara Green's functions since from now on we will only discuss this finite temperature formalism. 

The discussion above shows that at finite temperature one can get grand canonical ensemble averages of one-body operators using the Matsubara formalism. However, let us ask one more question: Can we get a system's static thermodynamic variables such as electronic Gibbs or Helmholtz energy, internal energy, and electronic entropy from dynamic (frequency dependent) variables such as Green's function and self-energy? 

The thermodynamics of a system  can be described by a thermodynamical potential, such as the grand potential~$\Omega$ \cite{vanleeuwentext}
\begin{equation}\label{LW_functional}
\Omega = \frac{1}{\beta}\{ \Phi -{\rm Tr}[\Sigma G+{\rm ln}(\Sigma-G^{-1}_0)]\},
\end{equation}
where the self-energy $\Sigma=\Sigma(i\omega_n)$ describes all the frequency dependent correlational effects present in the system and $G_0=G_0(i\omega_n)$ is the reference (usually non-interacting) system's Green's function and $G=G(i\omega_n)$ is the interacting Green's function. The interacting and non-interacting Green's functions are connected through the Dyson equation, $\Sigma=G^{-1}_0-G^{-1}$.
The $\Phi$ functional from Eq.~\ref{LW_functional} is called the Luttinger-Ward functional\cite{LW1960} \cite{footnote2} and is defined as 


\begin{equation}
\Phi= \sum_{m=1}^{\infty} \frac{1}{2m}{\rm Tr} [\sum_{n}\Sigma^{(m)}(i\omega_n) G(i\omega_n) ],
\end{equation}
where $\Sigma^{(m)}(i\omega_n)$ is a self-energy containing all irreducible and topologically inequivalent diagrams of order $m$.
The detailed derivation of Eq.~\ref{LW_functional} is presented in Ref.~\onlinecite{vanleeuwentext}.

Thus, the computational object that provides a connection between the static and dynamic quantities is  the Luttinger-Ward functional. This scalar functional  $\Phi=\hat{\Phi}[G]$ depends on the Green's function and has multiple important properties for Green's function theory.

\section{Properties of the Luttinger-Ward Functional}\label{sec_phi}
The formal properties of Luttinger-Ward functional have been discussed extensively before.  The functional has previously been applied to calculate energies for atoms and molecules \cite{van2006total,dahlen2006variational}, the electron gas \cite{VE}, as well as for the Hubbard lattice \cite{potthoff2003Hubbard, Janis, Janis2, potthoff2}. In the following section, we will only outline some of the most salient properties of the Luttinger-Ward functional {\color{blue} to} benefit the reader.

\subsection{Self-energy as a functional derivative}

The self-energy $\Sigma=\Sigma(i\omega_n)$ can be obtained as a functional derivative of the Luttinger-Ward functional
\begin{equation}
\beta \frac{\delta \hat{\Phi}[G]}{\delta G} = \hat{\Sigma}[G].
\end{equation}
Here, the self-energy is defined as a functional of Green's function that is evaluated independent of the Dyson equation. 
Consequently, in the non-interacting limit, where $\Sigma=0$, it follows that the Luttinger-Ward functional is zero itself, $\hat{\Phi}[G]=0$.

\subsection{Connection with the grand potential $\Omega$}

The grand potential is a number  but the mathematical object defined in Eq.~\ref{LW_functional} can be viewed more generally as a functional of Green's functions $\Omega[G]$.
When a Green's function is a self-consistent solution of the Dyson equation then the functional derivative $\frac{\delta \Omega[G]}{\delta G}=0$  since $\delta \Phi={\rm Tr}[\Sigma \delta G]$ and 
$\delta \Omega[G]= \frac{1}{\beta} \{ \delta \Phi  - {\rm Tr}[\delta \Sigma G   + \Sigma \delta G  - G \delta \Sigma]  \}$.
Consequently, we can conclude that the functional $\Omega[G]$ at the stationary point is equal to the grand potential. 
Having grand potential one gains access to the partition function (Z) since
\begin{equation}\label{eqn:gc_pt}
\Omega=-\frac{1}{\beta}{\rm ln} Z.
\end{equation}
The Helmholtz energy, $A=E-TS$, where $E$ is the internal energy and $S$ is the entropy of a system at a given temperature $T$, is connected to the grand potential as $A=\Omega+\mu N$, where $N$ is the number of electrons in the system.  
Thus, knowing the grand potential, we can easily calculate the 
electronic entropy as 
\begin{equation}
S=\frac{E-\Omega-\mu N}{T}
\end{equation}
as long as we have access to the internal energy of a system. The internal energy at a given temperature $T$ can be evaluated using
the Galitskii-Migdal formula
\begin{equation}
E=\frac{1}{2}{\rm {Tr}}\left[\left(h+F\right)\gamma\right]+\frac{2}{\beta}\sum_{n}^{N_\omega}{\rm {Re}\big(\rm Tr}[G(i\omega_n)\Sigma(i\omega_n)]\big)\label{eq:Etot},
\end{equation}
where $\gamma$ is the one-body density matrix, $h$ is the one-body Hamiltonian, $F$ is the Fock matrix of a system, and $N_\omega$ is the size of the imaginary grid.
Consequently, having access to the Luttinger-Ward functional of a system yields multiple electronic thermodynamic quantities.

\subsection{Universality}
Given two systems $A$ and $B$ at the same physical temperature $T$ and the same chemical potential $\mu$ described by two  Hamiltonians $\hat{H}_A=\sum_{ij}t^A_{ij}a_{i}^{\dagger}a_{j}+\sum_{ijkl}v_{ijkl}a_{i}^{\dagger}a_{j}^{\dagger}a_{l}a_{k}$ and $\hat{H}_B=\sum_{ij}t^B_{ij}a_{i}^{\dagger}a_{j}+\sum_{ijkl}v_{ijkl}a_{i}^{\dagger}a_{j}^{\dagger}a_{l}a_{k}$ such that they have the same two-body integrals $v_{ijkl}$ but different one-body integrals $t^{A}\ne t^{b}$ the Luttinger-Ward functional is same (universal) for both of them. Since $\hat{\Phi}^A=\hat{\Phi}^B$, then it must also hold that $\hat{\Sigma}^A(G)=\hat{\Sigma}^B(G)$. In other words, two systems described by different $G_0$ but having the same two-body interactions are described by the same Luttinger-Ward functional. 

\section{Evaluation of Luttinger-Ward Functional Within GF2}\label{sec_eval_phi_gf2}
\subsection{Description of the GF2 algorithm}
The self-energy, which we evaluate self-consistently in this work, is computed perturbatively at the second-order (GF2) level. GF2 is advantageous for many reasons, namely, among others it behaves qualitatively correct for moderately strongly correlated systems~\cite{GF2}, unlike methods such as MP2 or CCSD which tend to diverge in these cases. GF2 has small fractional charge and fractional spin errors \cite{fractional}. GF2 is carried out self-consistently on imaginary time $\tau$ and imaginary frequency $i\omega_n$ axes with a computational scaling of $\mathcal{O} (n_\tau N^5$) for molecular cases, where $n_\tau$ is a prefactor that depends on the size of the imaginary time grid and  $N$ is the number of orbitals present in the problem. We build the Green's function using the following expression
\begin{equation}
{G}(i\omega_n)=[(i\omega_n + \mu){S} - {F} - {\Sigma}(i\omega_n)]^{-1}
\end{equation}
where ${F}$ and ${S}$ are the Fock and overlap matrices in the atomic orbital (AO) basis, respectively, and $\mu$ is the chemical potential, which guarantees a correct particle number. To obtain $\Sigma(i\omega_n)$, we solve the Dyson equation given as
\begin{equation}
{\Sigma}(i\omega_n) = {G_0}(i\omega_n)^{-1} - {G}(i\omega_n)^{-1}
\end{equation}
where ${G_0}(i\omega_n)=[(i\omega_n + \mu){S} - {F}]^{-1}$ is the non-interacting Green's function while $G(i\omega_n)$ is the interacting Green's function since it contains the self-energy.
 To reduce the number of necessary  grid points, we employ a spline interpolation method to evaluate the Green's function \cite{spline} and Legendre orthogonal polynomials to expand the $\Sigma(\tau)$ matrix~\cite{legendre}. As pointed out previously, this self-consistent evaluation guarantees that both the Galitskii-Migdal and Luttinger-Ward energies are stationary with respect to the Green's function. For a full discussion of the algorithm and implementation details of GF2, we refer the reader to Refs.~\onlinecite{GF2,fractional}.
The main computational object in our evaluation is the self-energy in the AO basis, which can be expressed as 
\begin{equation}\label{se}
\begin{split}
\Sigma_{ij}(\tau) = -\sum_{klmnpq}^{ }G_{kl}(\tau)G_{mn}(\tau)G_{pq}(\tau) \times \\
\times v_{ikmq}(2v_{ljpn} - v_{pjln}).
\end{split}
\end{equation}

\noindent where $v_{ijkl}$ are the two-electron integrals in the AO basis in the chemist's notation.
Eq.~\ref{se} is evaluated in a Legendre polynomial basis \cite{legendre} to accelerate the calculation.

The details of a periodic GF2 implementation have been reported in Ref.~\cite{rusakov}. The basic differences from the molecular version include
solving the Dyson equation in $\textbf{k}$-space via 
\begin{equation}\label{GFk}
{G}^{k}(i\omega_n) = \left[(i\omega_n+\mu){S}^{k} - {F}^{k} - {\Sigma}^{k}(i\omega_n)\right]^{-1}
\end{equation}
and complicating the real-space self-energy, ${\Sigma}^{\mathbf{0g}}(\tau)$, evaluation by additional cell index summations:
\begin{equation}\label{GF2_pbc_real}
\begin{split}
\Sigma_{ij}^{\mathbf{0g}}(\tau) = -\sum_{\mathbf{g_1,\ldots,g_6}}\sum_{klmnpq}G_{k~l}^{\mathbf{g_3g_6}}(\tau)G_{m~n}^{\mathbf{g_1g_4}}(\tau)G_{p~q}^{\mathbf{g_5g_2}}(-\tau) \times \\ \times v_{i~m~q~k}^{\mathbf{0g_1g_2g_3}}(2v_{j~n~p~l}^{\mathbf{gg_4g_5g_6}}-v_{j~l~p~n}^{\mathbf{gg_6g_5g_4}}).
\end{split}
\end{equation}
${\Sigma}^{k}(i\omega_n)$ in Eq.~\ref{GFk} and ${\Sigma}^{\mathbf{0g}}(\tau)$ in Eq.~\ref{GF2_pbc_real} are interconvertible 
via corresponding Fourier transforms from the $\mathbf k$- to real-space and between the imaginary time and imaginary frequency domains also via Fourier transform.
The self-energy calculation according to Eq.~\ref{GF2_pbc_real} results in $\mathcal{O}(N^5N_{cell}^4n_{\tau})$ formal scaling of the 
computation cost with $N$ the number of orbitals in the unit cell and $N_{cell}$ the number of real space cells. This is typically a computational bottleneck of the GF2 self-consistency procedure.

The level of self-consistency at which the correlated Green's function equation should be iterated can depend on particular phases present in the phase diagram, such as Mott (see Ref.~\cite{Kotliar}). In particular, the  non-interacting Green's function $G_{0}(i\omega_n)$ build using updated Fock matrix or the correlated Green's function $G(i \omega_n)$ can re-enter the evaluation of the self-energy. We observe that the use of $G_{0}(i \omega_n)$ with the updated Fock matrix in the self-consistent evaluation of the self-energy is a well-behaved procedure when the strong correlations and the Mott phases emerge. The full self-consistent cycle, with $G(i \omega_n)$ re-entering the evaluation of the self-energy becomes ill behaved for these cases and we experienced difficulty converging it. We therefore use the former ``partial'' self-consistency ($G_{0}(i \omega_n)$ with the updated Fock matrix in the self-consistent evaluation of the self-energy) for the Mott regime. Such scheme is not uncommon in DMFT type calculations~\cite{Kotliar}. 


The expression for the total energy likewise acquires a cell summation according to Eqs. 13 and 14 in Ref.~\cite{rusakov}:
\begin{eqnarray}
&E_{tot} =E_{1b}+E_{2b}\\ \nonumber
&=\frac{1}{2} \sum_{\mathbf{g}, i, j}\gamma^{\mathbf{0g}}_{ij}(2h^{\mathbf{0g}}_{ij}+\left[\Sigma_{ \infty}\right]^{\mathbf{0g}}_{ij}) + \\
&+ \frac{2}{\beta}\sum_{\mathbf{g}, i, j}\operatorname{Re}\left[\sum_{n} G^{\mathbf{0g}}_{ij}(i\omega_n)\Sigma^{\mathbf{0g}}_{ij}(i\omega_n) \right]. \nonumber
\end{eqnarray}

\subsection{Evaluation of the grand potential in the $\mathbf k$-space}

The evaluation of  the Luttinger-Ward functional in conserving approximations \cite{baymkadanoff} such as the self-consistent second-order Green's function (GF2)\cite{GF2,fractional,legendre} is quite straightforward and was originally derived by Luttinger and Ward~\cite{LW1960} as

\begin{equation}
\Phi^{(2)} = \frac{1}{4}{\rm Tr}[\sum_{n}\Sigma^{(2)}(i\omega_n)G(i\omega_n) ],
\end{equation}

\noindent where $\Sigma^{(2)}(i\omega_n)$ is the frequency dependent part of the second-order self-energy. Since both $G(i\omega_n)$ and $\Sigma^{(2)}(i\omega_n)$ are readily available from a GF2 calculation, we are able to easily evaluate all terms of the functional.

The expression for grand potential from Eq.~\ref{LW_functional} can be conveniently reformulated as
\begin{equation}
\begin{split}
\Omega_{LW} = \frac{1}{2} {\rm Tr}[\gamma \Sigma_{\infty}] + {\rm Tr}[G \Sigma] 
\\ +{\rm Tr}[{\rm ln} \{ 1 - G \Sigma \} ]  + {\rm Tr}[{\rm ln} \{ G_0^{-1} \}].
\end{split}
\label{eqn:LW_reform}
\end{equation}

An excellent derivation of the above equation is given in Ref.~\onlinecite{dahlen2006variational} for molecular systems. However, as we mentioned before, for molecular systems the changes due to the electronic contributions of the Gibbs or Helmholtz energy are negligible due to the size of the gap. 

Here, we list detailed steps that need to be executed when dealing with crystalline systems where the electronic effects influencing Helmholtz energy can be significant.
In the periodic implementation, we evaluate the grand potential per unit cell, $\Omega^{\mathbf{00}}$.  The overall expression is given as a sum of all the components
\begin{equation}
\begin{split}
\Omega^{\mathbf{00}} = \Omega_{ \rm Tr[G \Sigma]}^{\mathbf{00}} + \Omega_{ \rm Tr[\rm ln \{ 1-G \Sigma \} ]}^{\mathbf{00}}
\\  + \Omega_{ \rm Tr[\rm ln \{ G_0^{-1} \} ]}^{\mathbf{00}} + \Omega_{\frac{1}{2} \rm Tr[\gamma \Sigma_{\infty}]}^{\mathbf{00}}
\end{split}
\label{eqn:LW_central}
\end{equation}

\noindent where we sum the contribution per unit cell for each term present in Eq.~\ref{eqn:LW_reform}. Due to the crystalline symmetry, it is often more convenient to calculate these quantities in $\mathbf k$-space and transform the resulting quantity to the real space rather than using an explicit real space representation of all the quantities involved.  All terms in Eq.~\ref{eqn:LW_central} containing the Green's function $G$, or self-energy $\Sigma$ were computed in $\mathbf k$-space and then Fourier transformed to real space. 
For example, to calculate the $ \Omega_{ \rm Tr[G \Sigma]}^{\mathbf{00}}$  contribution, we can first evaluate the $\mathbf k$-dependent quantity

\begin{equation}
\Omega^k_{\rm Tr[G \Sigma]} = \frac{2}{\beta} \sum_{i,j,n}^{N_{\omega}} G^k(i\omega_n)_{ij} \Sigma^k(i\omega_n)_{ji}
\label{eqn:Trace}
\end{equation}

\noindent where the indices $i$ and $j$ run over all atomic orbitals in a given $k-$block. Subsequently, we perform Fourier transform of  $\Omega^k_{\rm Tr[G \Sigma]}$ to yield the real space $\Omega^{\mathbf{00}}_{\rm Tr[G \Sigma]}$ contribution.

The most cumbersome evaluation is of the term $ \Omega_{ \rm Tr[\rm ln \{ 1-G \Sigma \} ]}^{\mathbf{00}} $, which requires diagonalization of the matrix $ G^k \Sigma^k + (G^k \Sigma^k)^{\dagger} -  G^k \Sigma^k (G^k \Sigma^k)^{\dagger}$, where the matrices $G^k$ and $\Sigma^k$ are understood to be dependent on $i\omega_n$. Once the eigenvalues of this matrix, $\epsilon^{k}_{i}$, have been computed for each imaginary frequency point, $i\omega_n$, we have

\begin{equation}
 \Omega_{ \rm Tr[\rm ln \{ 1-G \Sigma \} ]}^{k} = \frac{2}{ \beta } \sum_{n, i}^{N_{\omega}} \rm ln \{ 1 - \epsilon^{k}_{i} \},
\label{eqn:logterm}
\end{equation}

\noindent which can be Fourier transformed to the real space. 

To include the contribution from the Fock matrix, $\Omega_{ \rm Tr[\rm ln \{ G_0^{-1} \} ]}^{\mathbf {00}}$, we calculate 

\begin{equation}
\Omega_{ \rm Tr[\rm ln \{ G_0^{-1} \} ]}^{k}=\begin{cases}
   \frac{1}{\beta} \: \sum_{i}\rm  ln(1 + e^{\beta(\epsilon^{k}_{i} - \mu)}) + \epsilon^{k}_{i}, & \text{if $\epsilon^{k}_{i}-\mu <0$}\\
       \frac{1}{\beta} \: \sum_{i}\rm ln(1 + e^{-\beta(\epsilon^{k}_{i} - \mu)}), & \text{otherwise}.
\end{cases}
\label{eqn: Fock}
\end{equation}

\noindent where by analyzing if the term $\epsilon^{k}_{i}-\mu$ is smaller or greater than zero we account for the cases where the absolute value of the Fock matrix eigenvalue can be large leading to numerical problems if only one branch of the above expression is used.
This term accounts for occupation changes with temperature. For high $\beta$ values (low values of the actual temperature $T$), the expression reduces to 

\begin{equation}
\Omega_{ \rm Tr[\rm ln \{ G_0^{-1} \} ]}^{k}=\sum_{N_{occ}}\epsilon^{k}_{i}, 
\end{equation}

\noindent which allows electrons to occupy only the lowest available state, as is expected at a very low temperature. 

We calculate the term $\Omega_{\frac{1}{2} \rm Tr[\gamma \Sigma_{\infty}]}^{\mathbf{00}}$ directly in the real space as 
$\frac{1}{2} {\rm Tr}[\gamma \Sigma_\infty]$, since in the AO basis, the decay of both $\gamma$ which is the density matrix and $\Sigma_{\infty}$ which is the frequency independent part of the self-energy is rapid enough for a relatively few
number of cells to assure a converged value of $\Omega_{\frac{1}{2} \rm Tr[\gamma \Sigma_{\infty}]}^{\mathbf{00}}$.

\section{Results} \label{sec_results}
\subsection{HF molecule}
In this subsection, for a simple molecular example,  a hydrogen fluoride molecule, we provide a calibration of the thermodynamic quantities such as internal energy $E$, Helmholtz energy $A$, and entropy $S$, which are evaluated at the GF2 level and compared to the full configuration interaction (FCI) calculation. The evaluation of the thermodynamic quantities at the FCI level can be done only for very small molecular examples,  such as HF,  since such a system has only 10 electrons and 6 basis functions in the STO-3G basis, resulting in a small number of possible configurations necessary to evaluate the FCI grand potential. Note that to describe a true physical system at very high temperature we would require a very large basis set. Here, we use HF as a model molecular system that is calculated in a minimal basis set solely to enable comparison of GF2 with FCI. The full configuration interaction (FCI) quantities for HF molecule were calculated previously by Kou and Hirata~\cite{FCISoHirata}. In addition, we show the same system calculated with finite-temperature at the Hartree-Fock level. We would like to emphasize that we provide this molecular example as a benchmark only, in order to compare our method with highly accurate FCI quantum chemical data. Typically electronic contributions to thermodynamics are not considered for molecular systems.

Let us first note that in order to calculate the FCI partition function and subsequently grand potential in the grand canonical ensemble, we need to evaluate 
\begin{equation}
Z^{GC}=\sum_{N=0}^{2n}\sum_{S_z}\sum_{i}\langle \Phi_{i}^{(N,S_z)}|exp\{-\beta(\hat{H}-\mu\hat{N})\}| \Phi_{i}^{(N,S_z)}\rangle,
\end{equation}
where the $\Phi_{i}^{(N,S_z)}$ is the FCI wave function with $N$ electrons and $S_z$ quantum number and the number of possible occupation runs from 0 to $2n$, where $n$ is the number of orbitals. Consequently, we need to explicitly obtain the information about every possible excited state present in the system with different number of electrons. Such a task quickly becomes impossible for any larger systems. In contrast, in Green's function methods, we never need to explicitly evaluate any information concerning specific excited states. It is sufficient to evaluate Eq.~\ref{eqn:LW_reform} and then Eq.~\ref{eqn:gc_pt} to obtain the grand canonical partition function. Thus, even for relatively large systems such calculations remain feasible.

Results from our calibration are shown in  Fig.~\ref{fig:FCI_GF2_HF}. The numerical values used in the plots are tabularized in the Supplemental Information. 
The detailed description of the grids on which we evaluate $\Sigma(\tau)$ and $G(i\omega_n)$ can be found in Ref.~\cite{footnote1}. 
For the hydrogen fluoride molecule the temperature range is huge due to the size of the Hartree-Fock HOMO-LUMO gap in this system, which is around 1.0 a.u. corresponding to a temperature of around $3.0 \e{5}$ K. Consequently, to make every state accessible to the electrons in this system, we require extremely high temperatures, as indicated by our results.

In the very high temperature limit, both finite temperature Hartee-Fock and GF2 yield the internal energy (E), Helmholtz energy (A),  and entropy (S) in excellent agreement with FCI results. 
This is of course expected since at very high temperatures the electronic behavior is well described by mean field theories.
For the intermediate temperatures, GF2 thermodynamic quantities (E, S, and A) are closer to FCI  than thermodynamic quantities obtained in finite temperature Hartee-Fock. For this system at intermediate temperatures, the GF2 thermodynamic quantities are always overestimated while the finite temperature Hartree-Fock always underestimate them in comparison to FCI.
The GF2 and Hartree-Fock entropies for low temperatures are well recovered and comparable. As expected, the low temperature GF2 internal energy is closer to FCI  than the one evaluated using finite temperature Hartree-Fock.
 For our very lowest temperature ($10^3$ K), we recover a small negative entropy. This is a numerical artifact brought about from the level of convergence of the internal energies (1.0\e{-5} a.u.) and the expression for entropy which requires multiplication by $\beta$, $S=\beta (E-\Omega-\mu N)$. Thus, the smallest error $\approx 10^{-5}$ in  the energy will result in $ \approx 10^{-3}$ error in the entropy due to multiplication by $\beta=100$. 

\begin{figure*}
\includegraphics[width=2.25in]{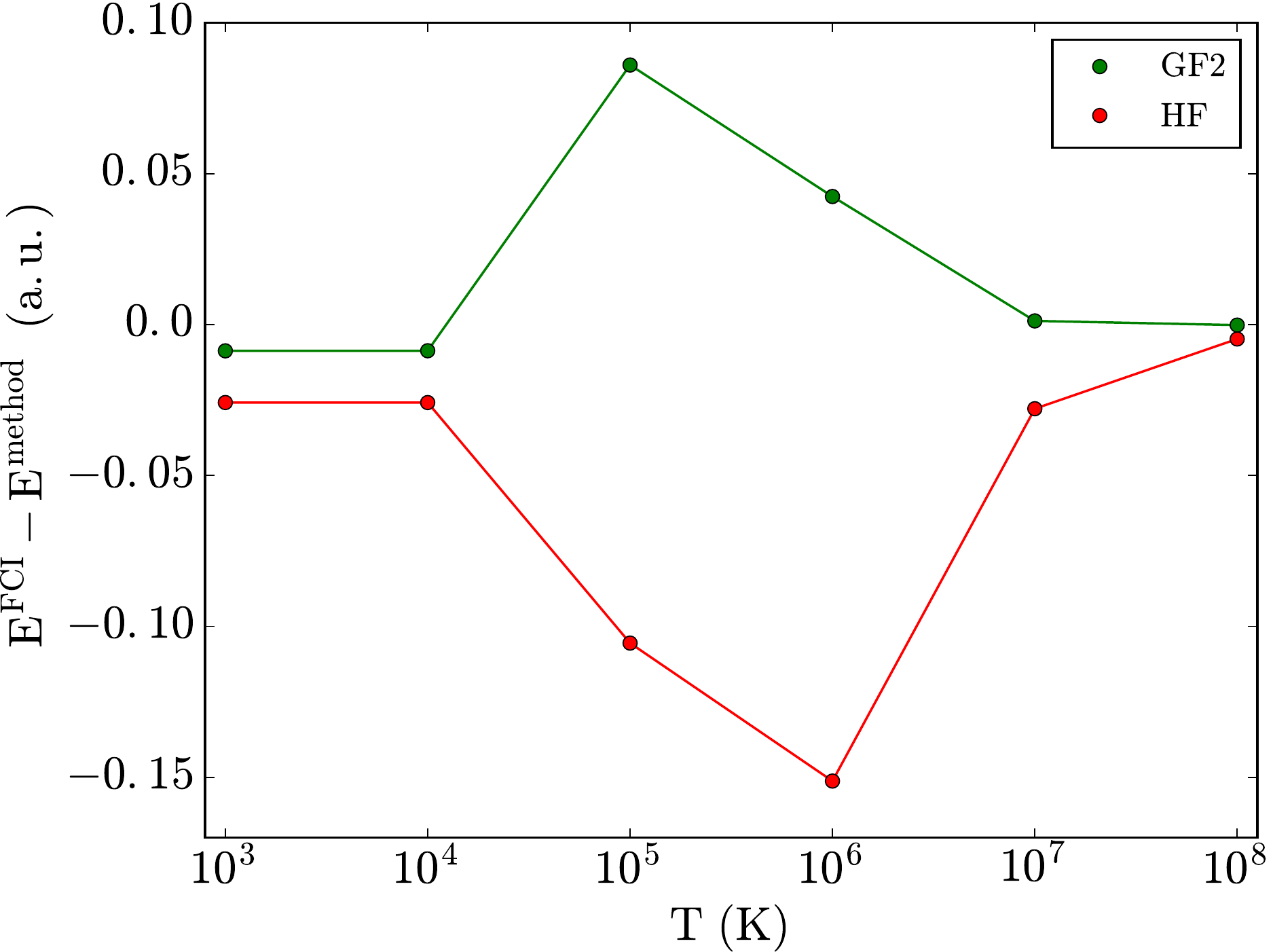}
\includegraphics[width=2.25in]{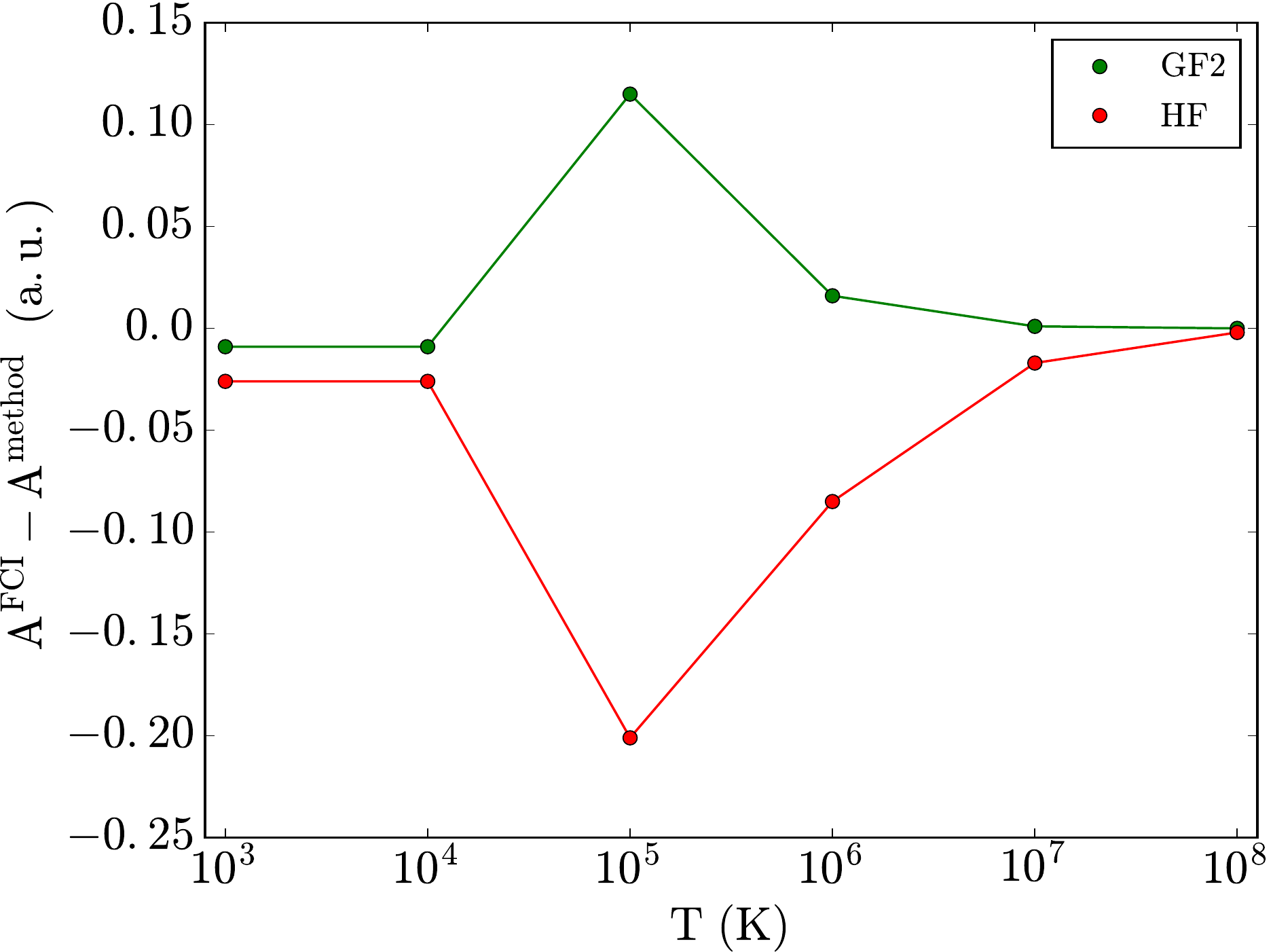}
\includegraphics[width=2.25in]{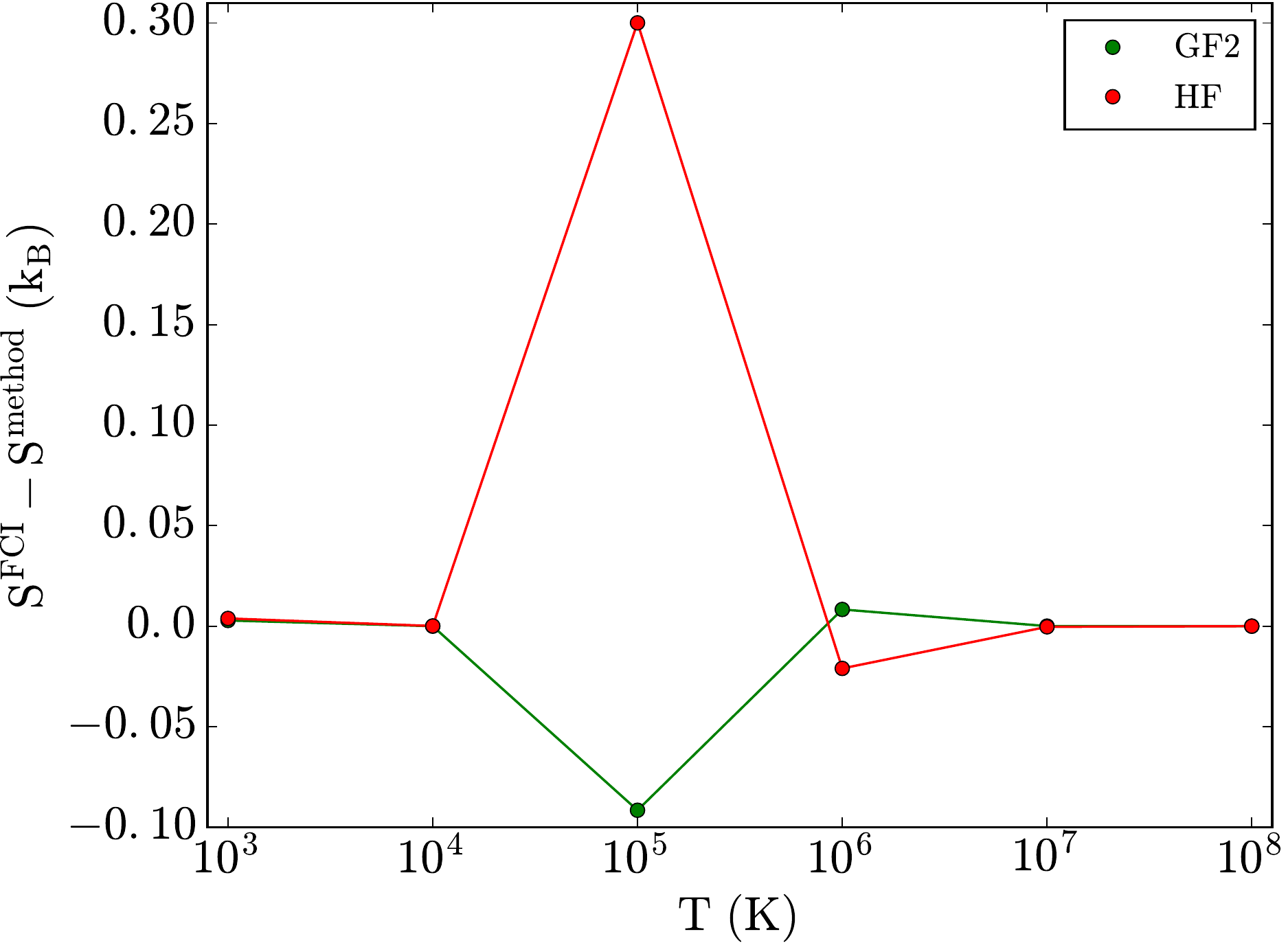}
\caption{The differences between GF2, finite temperature Hartree-Fock and FCI for the hydrogen fluoride molecule at various temperatures.}
\label{fig:FCI_GF2_HF}
\end{figure*}

\subsection{Periodic calculation of 1D hydrogen}

In our previous work, GF2 was implemented for periodic systems and applied to a 1D hydrogen solid~\cite{rusakov} in the mini-Huzinaga \cite{huzinaga1985basis} basis set. We consider the same system in this work, where we have used 5,000 Matsubara frequencies to discretize the Green's function in the frequency domain, 353 imaginary-time points, and 27 Legendre polynomials. We have found this grid size is sufficient to evaluate energy differences between the systems at various temperatures. Up to 73 real space unit cells appear in the self-energy evaluation (Eq.~\ref{GF2_pbc_real}). 

This system is simple enough to be a test bed for self-consistent Green's function theory; however, it displays a phase diagram that is characteristic of realistic solids. At different internuclear separations, corresponding to different pressures, we were able to recover multiple solutions. Although yielding different electronic energies and different spectra, these solutions can be mathematical artifacts of the nonlinear self-consistency procedure present in GF2; however, they can also have physical meaning corresponding to different solid phases. 

To decide which phase is more stable at a given temperature, it is necessary to consider the Helmholtz energies that we are able to obtain from the Luttinger-Ward functional for every solution. 
Previously, for the inverse temperature of $\beta=100$, at most of the geometry points, we have identified two possible phases with different internal energies, $E$, that were obtained starting the iterative GF2 procedure either from an insulating or metallic solution. The results of our investigation can be found in Fig.~4 of Ref.~\onlinecite{rusakov}. Currently, to analyze the stability of the solutions, for a range of inverse temperatures $\beta=25,75,100$, we discuss internal energy $E$, Helmholtz energy $A$, and the entropic contribution $TS$ to the Helmholtz energy. We also improved our convergence criteria not only converging the internal energy $E=E_{1b}+E_{2b}$ (as we have done in the previous work) but also converging both the $E_{1b}$, $E_{2b}$, and the Helmholtz energy separately. This much more stringent procedure to analyze convergence of GF2 leads us to slightly revised solutions for the 1D hydrogen solid which we discuss in the subsequent sections.

The spectra are produced from analytical continuation of the imaginary axis Green's function $G(\mathbf{k}, i\omega_n)$ to the real axis $G(\mathbf{k}, \omega_n)$ \cite{ALPS}. The spectral weight is proportional to $\text{Im}G(\mathbf{k}, \omega_n)$. A 2D color projection of the spectral function on the $(\mathbf{k},\omega_n)$ plane can be viewed as a ``correlated band structure'' analogous to the conventional band
structure within effective one-electron models such as Hartree--Fock and DFT. As in one-electron models, zero spectral
weight at the Fermi energy $\omega_F$ is indicative of a gapped system. 
The peaks emerging immediately below and above $\omega_F$ for a given $\mathbf{k}$ correspond to the energies of the highest occupied (HOCO) and lowest unoccupied (LUCO) crystalline orbitals, respectively. We should stress, however, that since $G(\mathbf{k}, \omega_n)$ is a many-body correlated
Green's function, such correspondence is not rigorous and merely serves as a convenient analogy.

\subsubsection{Short bond length/high pressure}
\begin{figure*}
\includegraphics[bb=0 0 1050 600, width=6.69in]{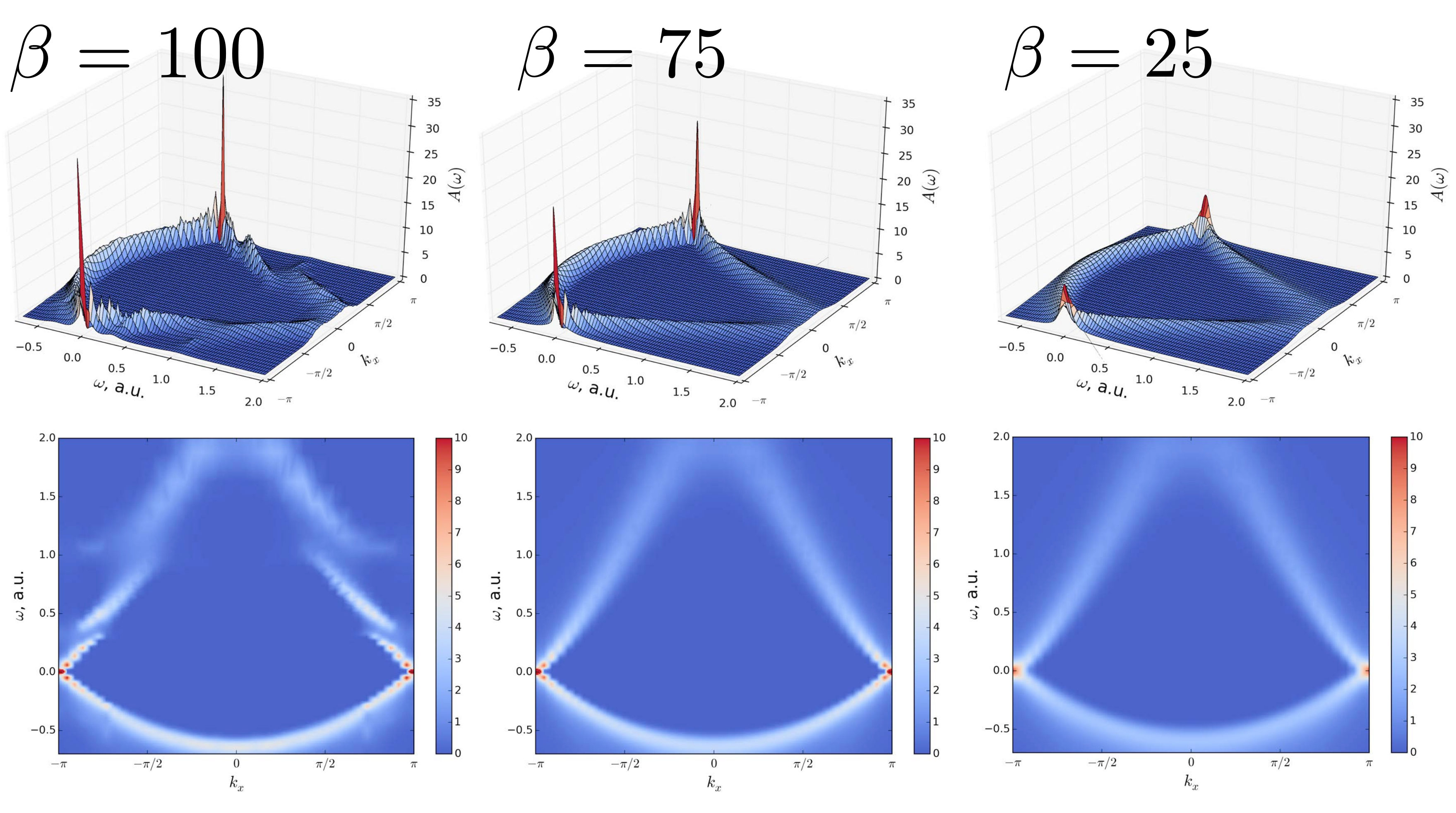}\\
\caption{ Spectral functions and projections at different inverse temperatures for the metallic solution of 1D periodic hydrogen solid at R=0.75 \AA.}
\label{fig:075}
\end{figure*}

At the interatomic separation of 0.75~\AA, \ we recovered only one gapless, metallic solution for all the values of inverse temperature ($\beta=25$, 75, 100). We established that starting from two different initial guesses leads in both cases to two final solutions that were different in internal energy and Helmoltz energy by less than $10^{-4}$ a.u. Consequently, we deemed that we obtained the same metallic solution in both cases. The spectral functions and spectral projections for this metallic solution at different values of inverse temperature are shown in Fig.~\ref{fig:075}. For all the temperatures examined in this short bond length regime, the self-energy displays a Fermi liquid character.

In our previous work~\cite{rusakov}, we observed a small internal energy difference between the solutions obtained using different starting point at $\beta=100$. We currently observe that this difference can be eliminated if we assure that the convergence criteria are not only fulfilled for the total energy $E_{tot}=E_{1b}+E_{2b}$ but also both the $E_{1b}$ and $E_{2b}$ components separately.

\subsubsection{Intermediate bond length/intermediate pressure}
\begin{figure*}
\includegraphics[bb=0 0 1050 600, width=6.69in]{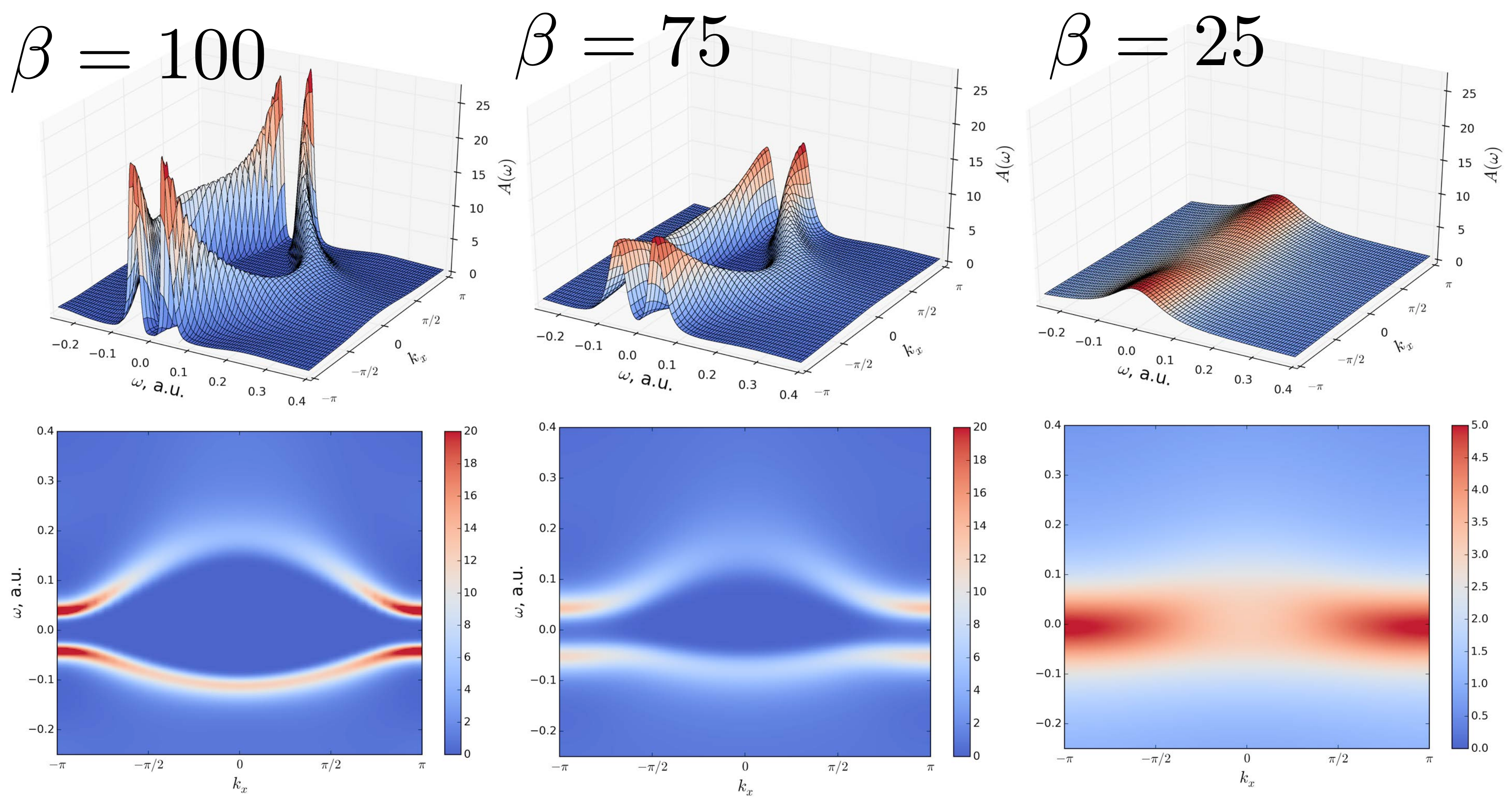}\\
\includegraphics[bb=0 0 1050 600, width=6.69in]{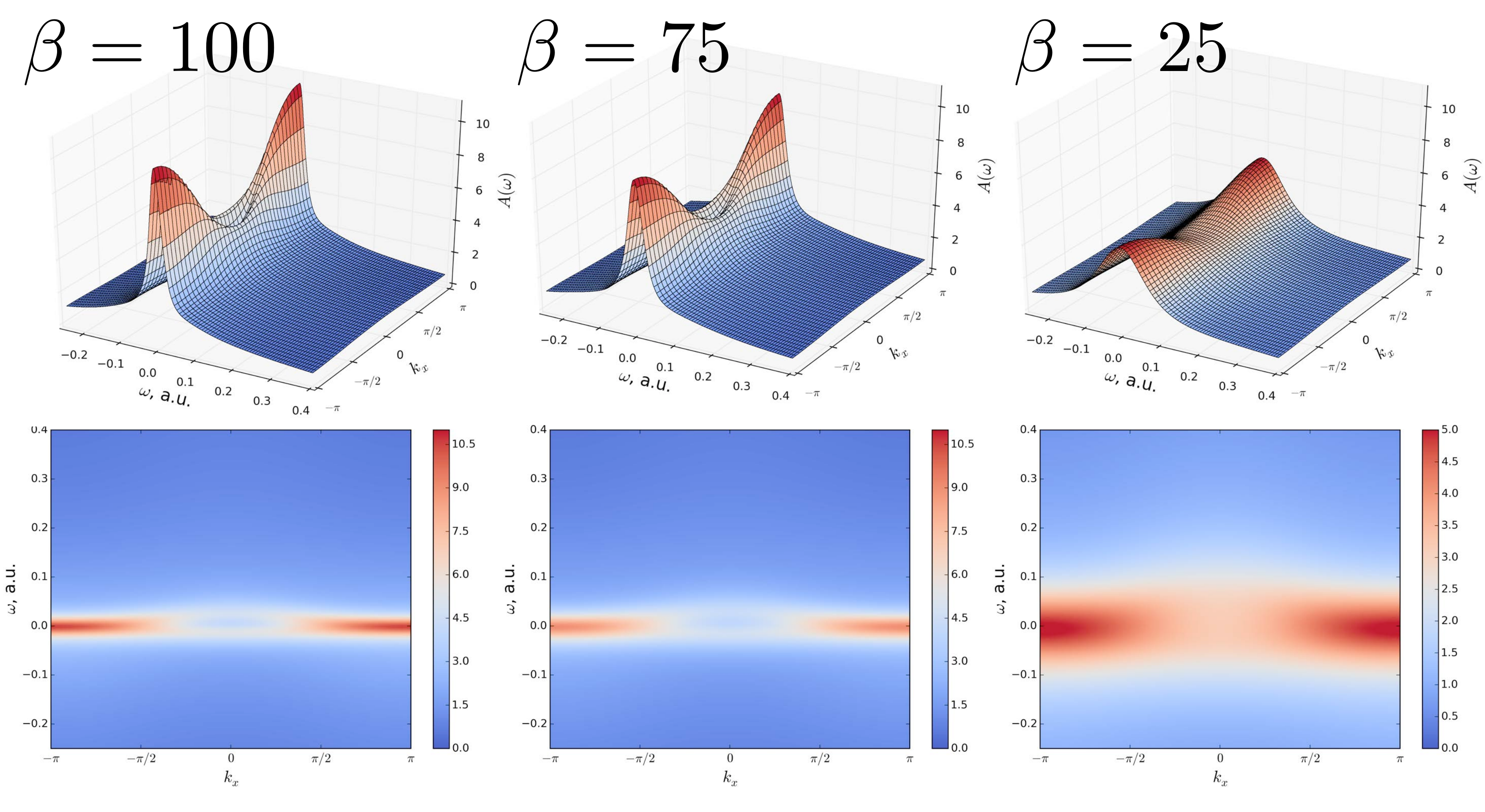}
\caption{Spectral functions and projections at different inverse temperatures for the band insulator (``solution 1") and the metallic solution (``solution 2") of 1D periodic hydrogen solid at R=1.75 \AA. Note the difference in scale between ``solution 1" and ``solution 2".}
\label{fig:175}
\end{figure*}

Spectral functions and projections for the 1D hydrogen solid with an interatomic separation of 1.75~\AA \ are presented in Fig.~\ref{fig:175}.  
We have displayed differences in internal energy (E), entropy (written as -T$\Delta$S), and Helmholtz energy (A) in Table ~\ref{tab:1.75data}.
For this system at the inverse temperatures of $\beta=100$ and $75$ we obtained two solutions from two different initial guesses. We are able to characterize the first solution (``solution 1") as a band insulator since the spectral function shows a gap and the self-energy displays a Fermi liquid profile. The second solution (``solution 2") is gapless and therefore a metal. At an inverse temperature of $\beta=25$, both ``solution 1" and ``solution 2" obtained from two different starting guesses have the same spectra and identical Helmholtz energy, indicating that there is only one stable solution --- a single phase. Note that in Fig.~\ref{fig:175} for $\beta=25$ both the spectral functions for ``solution 1" and ``solution 2" seem to have different heights; however, it is an illusion since both spectral functions are plotted with a different z-axis range.  We deem that the Helmholtz energy is identical for both these solutions at $\beta=25$ since the obtained differences are below our convergence threshold which is $1\times 10^{-5}$  a.u.

For two lower temperatures ($\beta=100$ and $75$) by comparing thermodynamic quantities, we are able to determine which phase, ``solution 1" or ``solution 2", is the most thermodynamically stable.  From the data in Table ~\ref{tab:1.75data}, we are able to determine that ``solution 1" is the most stable phase from the positive value of $\Delta$A at both $\beta$=100 and $\beta$=75. It is also interesting to note, that this solution is stable due to the entropic factor not due to the difference in the internal energy.
\begin{table}
\setlength{\tabcolsep}{8pt}
\renewcommand{\arraystretch}{1.6}
\begin{ruledtabular}
\begin{tabular}{c c c c  }
$\beta$ & $\Delta$ E & -T $\Delta$ S &  $\Delta$ A\\
\hline
100 & -0.00370 & 0.10854  & 0.10484 \\
75 & -0.00528 & 0.09971 & 0.09443\\
25 & 1.36\e{-8}  & -2.94\e{-6}  &  -2.93\e{-6}\\
\end{tabular}
\end{ruledtabular}
\caption{Thermodynamic data for a 1D periodic hydrogen solid with separation R=1.75 \AA .  The units for $\beta$ are 1/a.u. The units for all other quantities are a.u. All values are obtained by subtracting ``solution 1" from ``solution 2". $\Delta$ E =E$_{sol2}$-E$_{sol1}$, $\Delta$ A =A$_{sol2}$-A$_{sol1}$, $\Delta$ S =S$_{sol2}$-S$_{sol1}$. Note that the quantities at $\beta$=25 are below the precision of convergence (1 \e{-5}).}
\label{tab:1.75data}
\end{table}

\subsubsection{Long bond length/low pressure}
\begin{figure*}
\includegraphics[bb=0 0 1050 600, width=6.69in]{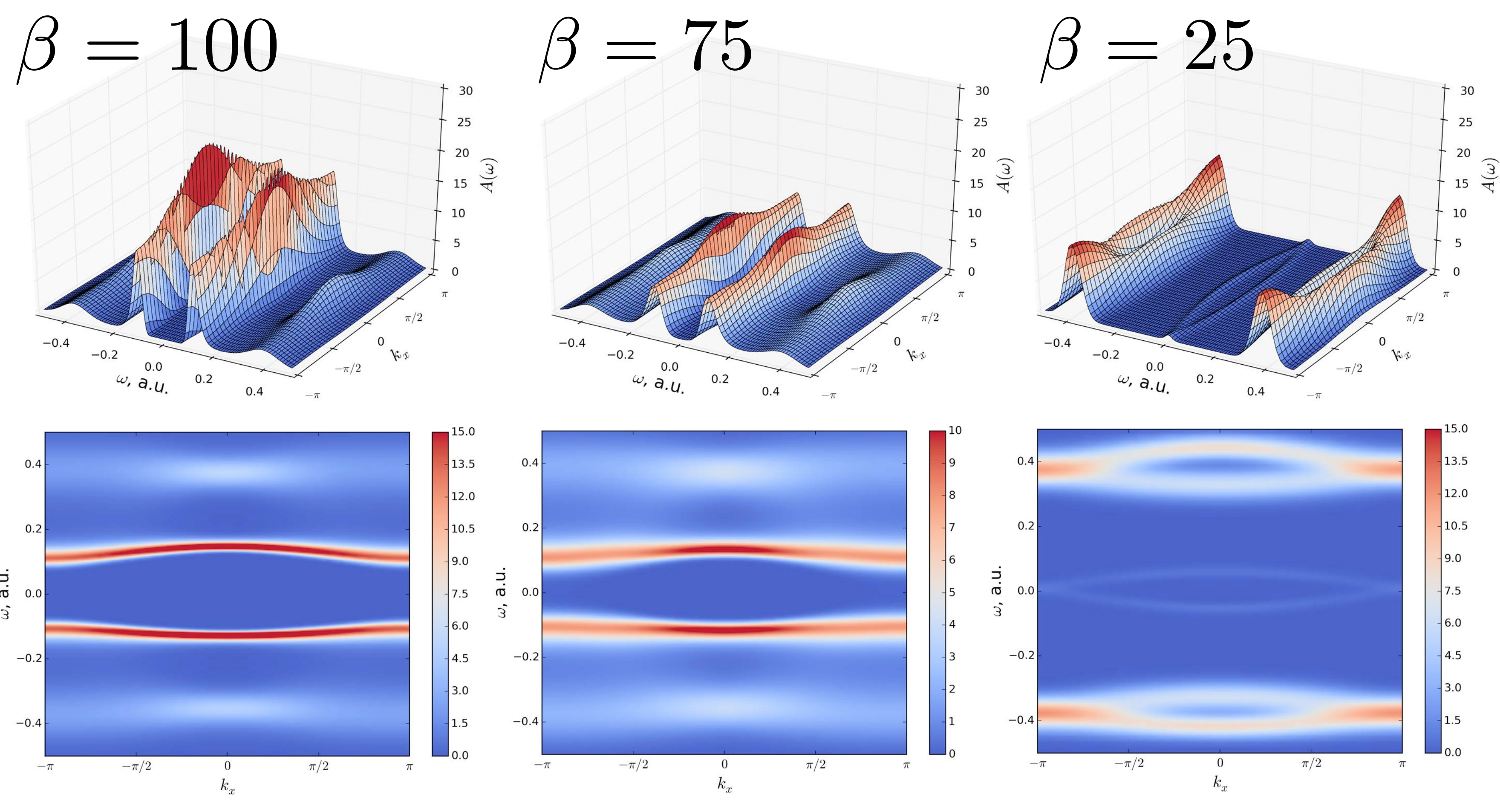} \\
\includegraphics[bb=0 0 1050 600, width=6.69in]{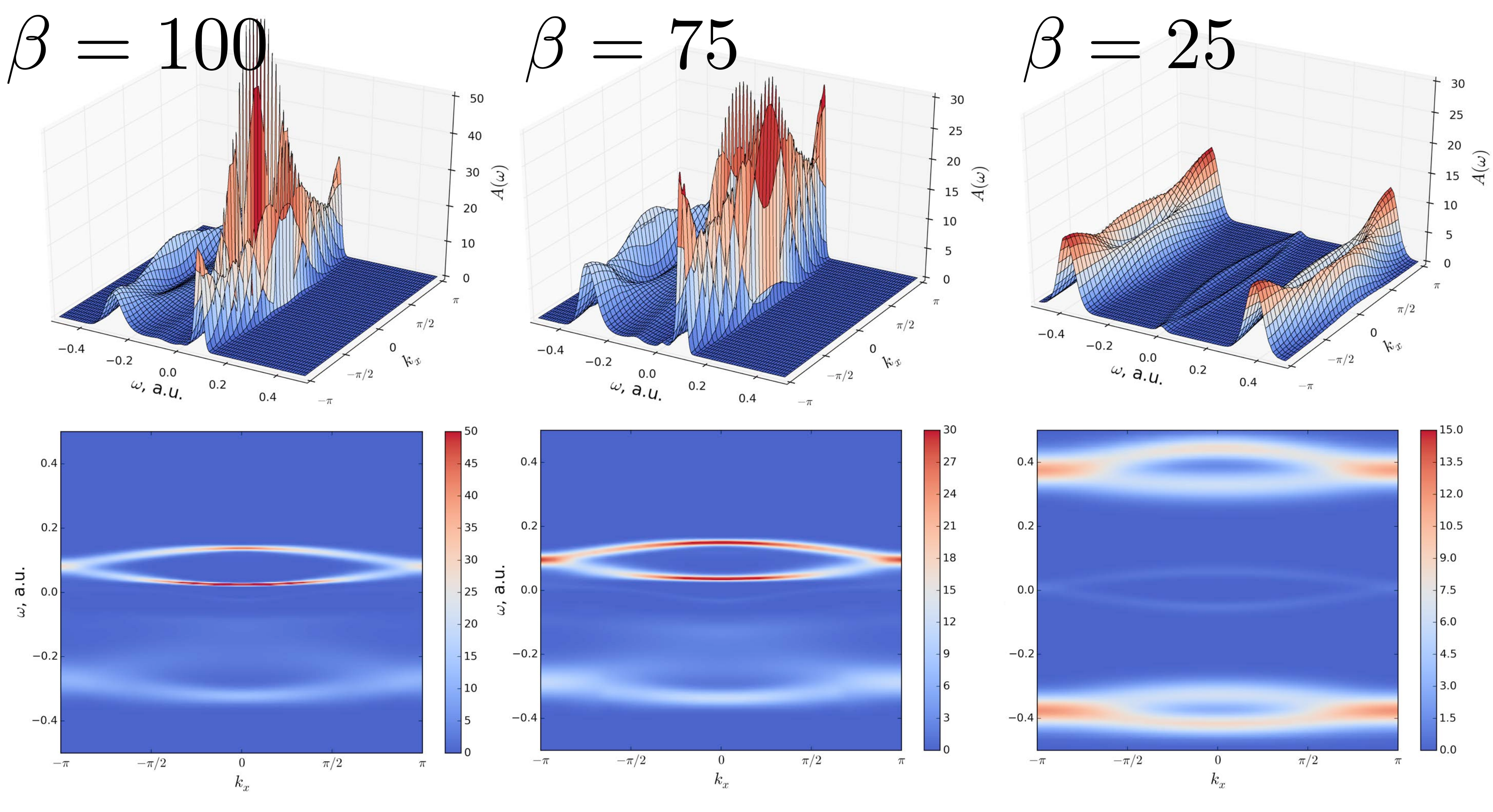}
\caption{Spectral functions and projections at different inverse temperatures for the band insulator (``solution 1") and the metallic solution (``solution 2") of 1D periodic hydrogen solid at R=2.5 \AA. }
\label{fig:2.5}
\end{figure*}

Similar to the R=1.75 \AA \ regime, for an interatomic separation of 2.5 \AA \  at inverse temperatures $\beta=100$ and $75$, we recover two solutions from the two different initial guesses, see Fig.~\ref{fig:2.5}. At a high temperature $\beta$=25 we observe only one solution independent of the initial guess, and thus only a single phase is present. From the positive value of $\Delta$A we are able to see that ``solution 1" is the most stable phase at $\beta$=75 and $\beta$=100 (Table ~\ref{tab:2.5data}). This solution is a band insulator since it has a Fermi liquid self-energy profile at these temperatures. 
The other solution, denoted as ``solution 2" at low temperature is metallic and changes into a Mott insulator at high temperatures. 
The internuclear distance of   R=2.5 \AA \ is close to the region where 
 the phase transition occurs, thus results obtained from GF2 which is a low order perturbation expansion may not be reliable. The low level perturbation theories such as GF2 are known to be more accurate deep within a phase and can experience problems near the phase transition point.

\begin{table}
\setlength{\tabcolsep}{8pt}
\renewcommand{\arraystretch}{1.6}
\begin{ruledtabular}
\begin{tabular}{c c c c  }
$\beta$ & $\Delta$ E & -T $\Delta$ S &  $\Delta$ A\\
\hline
100 & 0.08489  & 0.16257  & 0.24746  \\
75 &  0.07103& 0.16155 & 0.23259 \\
25 & 6.09\e{-8}  & 2.96\e{-5}  & 2.97 \e{-5}\\
\end{tabular}
\end{ruledtabular}
\caption{Thermodynamic data for a 1D periodic hydrogen solid with separation R=2.5 \AA .  The units for $\beta$ are 1/a.u. The units for all other quantities are a.u. All values are obtained by subtracting ``solution 1" from ``solution 2". $\Delta$ E =E$_{sol2}$-E$_{sol1}$, $\Delta$ A =A$_{sol2}$-A$_{sol1}$, $\Delta$ S =S$_{sol2}$-S$_{sol1}$. Note that the quantities at $\beta$=25 are below or comparable with the precision of convergence (1 \e{-5}).}
\label{tab:2.5data}
\end{table}



\begin{figure*}
\includegraphics[bb=0 0 1050 600, width=6.69in]{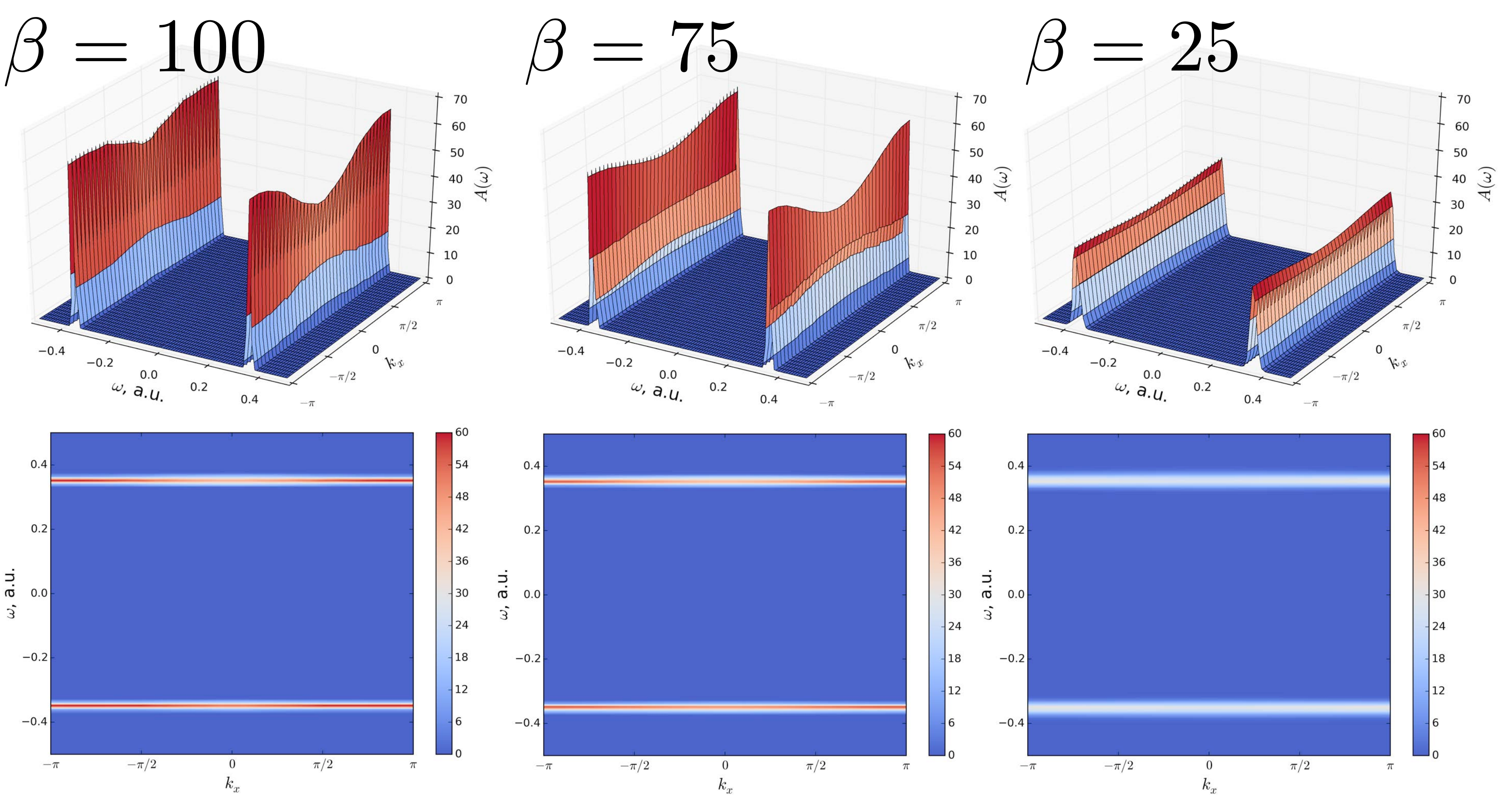} 
\caption{Spectral functions and projections at different inverse temperatures for the Mott insulator present in 1D periodic hydrogen solid at R=4.0 \AA.}
\label{fig:4.0}
\end{figure*}

Finally, at an interatomic separation of R=4.0 \AA , for all the temperatures, both initial guesses yield the same converged GF2 result - a single phase. The spectra as a function of inverse temperature can be seen in Fig.~\ref{fig:4.0}. This single solution is a Mott insulator as confirmed by the divergent imaginary part of the self-energy. 


\subsubsection{GF2 phase diagram for 1D hydrogen solid}

\begin{figure}
\includegraphics[bb=0 0 1050 800, width=3.375in]{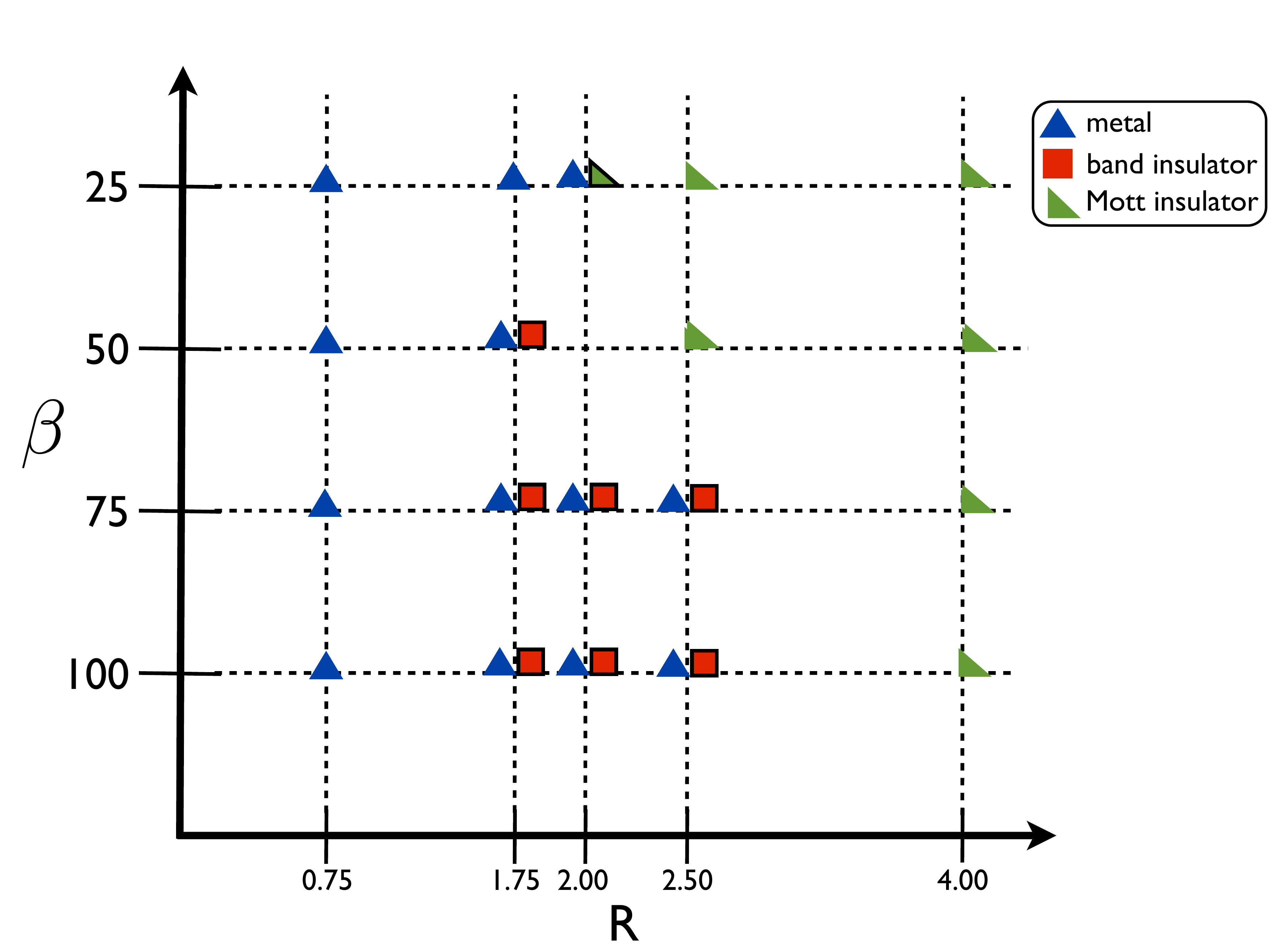}
\caption{A phase diagram containing all distances and temperatures calculated for a 1D hydrogen solid. Where multiple phases exist, the most stable phase is outlined in black. $\beta$ is inverse temperature in units 1/a.u. and R is the separation between hydrogens in \AA .}
\label{fig:phase}
\end{figure}

It is instructive now to collect all the data and construct a simple phase diagram for the 1D hydrogen solid as a function of inverse temperature $\beta$ and intermolecular distance R. We have plotted the phase diagram in Fig.~\ref{fig:phase}. 
On this diagram, for these regions where two phases coexist, we denoted the most stable phase according to the Helmholtz energy $A$ by framing its symbol using a black line.

For the shortest bond length (R=0.75 \AA ), the 1D hydrogen solid remains metallic at all temperatures considered and only one phase is present. 
At intermediate bond lengths (R=1.75, 2.0 \AA), multiple  phases coexist. At lower temperatures, we recover both a metal and band insulator as possible solutions, with the band insulator being the most stable phase according to the Helmholtz energies. This is in line with physical intuition that we should recover an insulator rather than a metal at low temperature.
For R=1.75 \AA, at temperatures higher than $\beta=50$, we recover only the metallic phase. 

For R=2.0 \AA \ at high temperature ($\beta=25$), we see both a metal and a Mott insulator coexisting, with the Mott insulator being the most stable phase. At lower temperatures, we see the coexistance of a metallic and band insulator solution. The Helmholtz energy indicates that band insulator is the most stable phase.

At a longer bond length of (R=2.5 \AA), multiple phases are still present. As in the cases of intermediate bond length, we recover both a band insulator and a metal solution, with Helmholtz energy favoring the band solution. We would like to reiterate that a phase transition occurs somewhere in this intermediate region, and it is likely that the results of a second-order perturbation theory may not be accurate enough. At higher temperatures, we recover a Mott insulator as the only phase present. 

For the largest separation (R=4.0 \AA), the system remains a Mott insulator at all studied temperatures. 

\subsection{Periodic calculation of 1D boron nitride}
In this section we present a periodic calculation of 1D boron nitride (BN) at inverse temperatures $\beta=$70, 75, and 100. The B--N distance is set to 1.445~\AA, the bond length in the corresponding 2D system \cite{C4CS00102H}. For these calculations we used a modification of the ANO-pVDZ Gaussian basis set~\cite{doi:10.1021/ct100396y} from which, in order to avoid linear dependencies, we removed the diffuse
(below 0.1) exponents and polarization functions. Larger grids of 30000 Matsubara frequencies and 100 Legendre polynomials were found necessary for an adequate discretization for the self-energy and Green's function. The self-energy is evaluated encompassing 39 cells --- as many as needed for a converged Hartree--Fock exchange.

\begin{figure}
\includegraphics[bb=0 0 1050 800,width=3.375in]{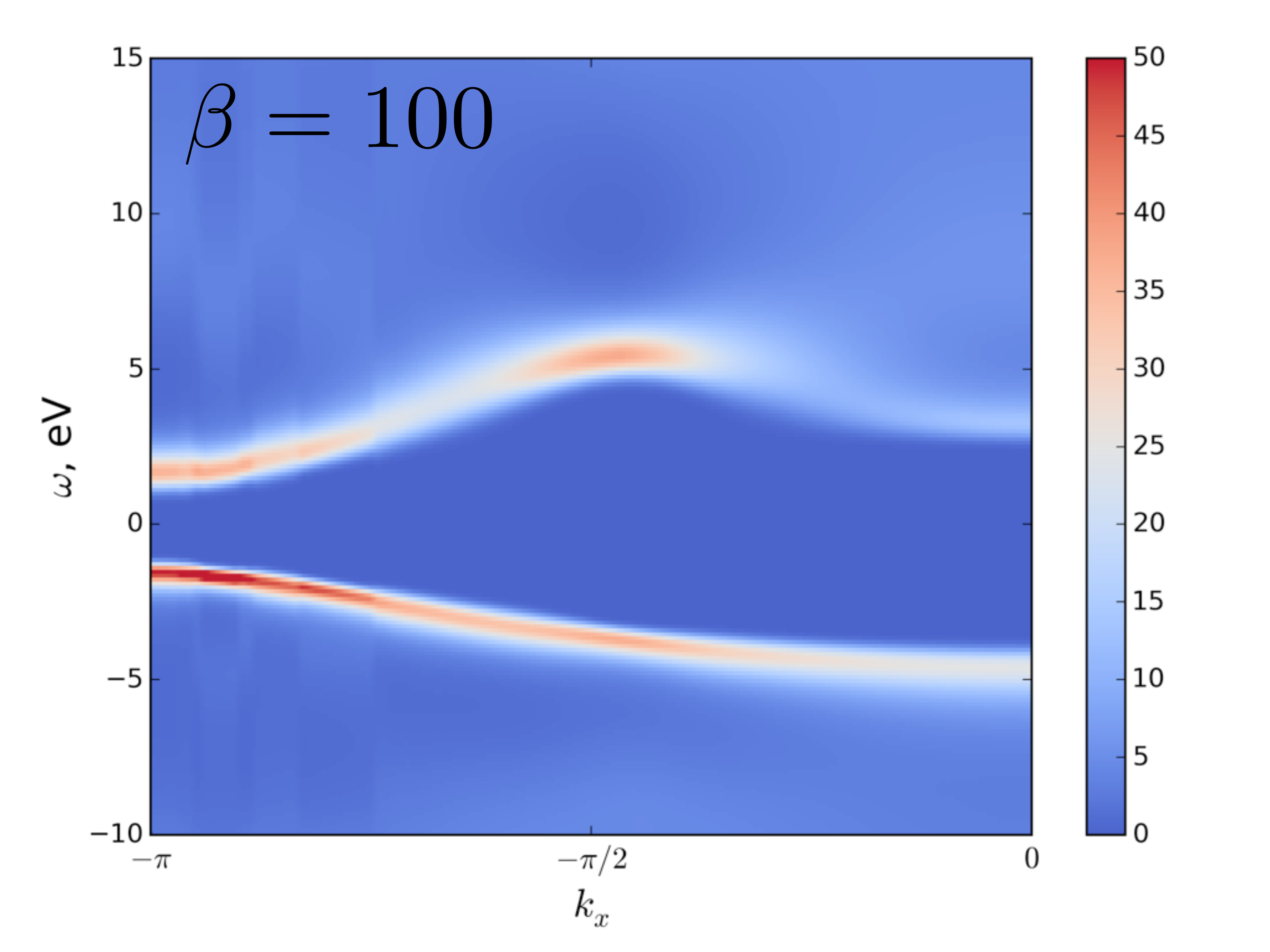}
\caption{Spectral projection in the vicinity of the Fermi energy at $\beta$=100 of periodic 1D boron nitride solid. Only the ``correlated bands'' corresponding to the conventional HOCO's and LUCO's (see Sec.~V.~B) are displayed. Note that the energy scale is in eV. }
\label{fig:BN_proj}
\end{figure}

\begin{figure}
\includegraphics[bb=0 0 1050 800,width=3.375in]{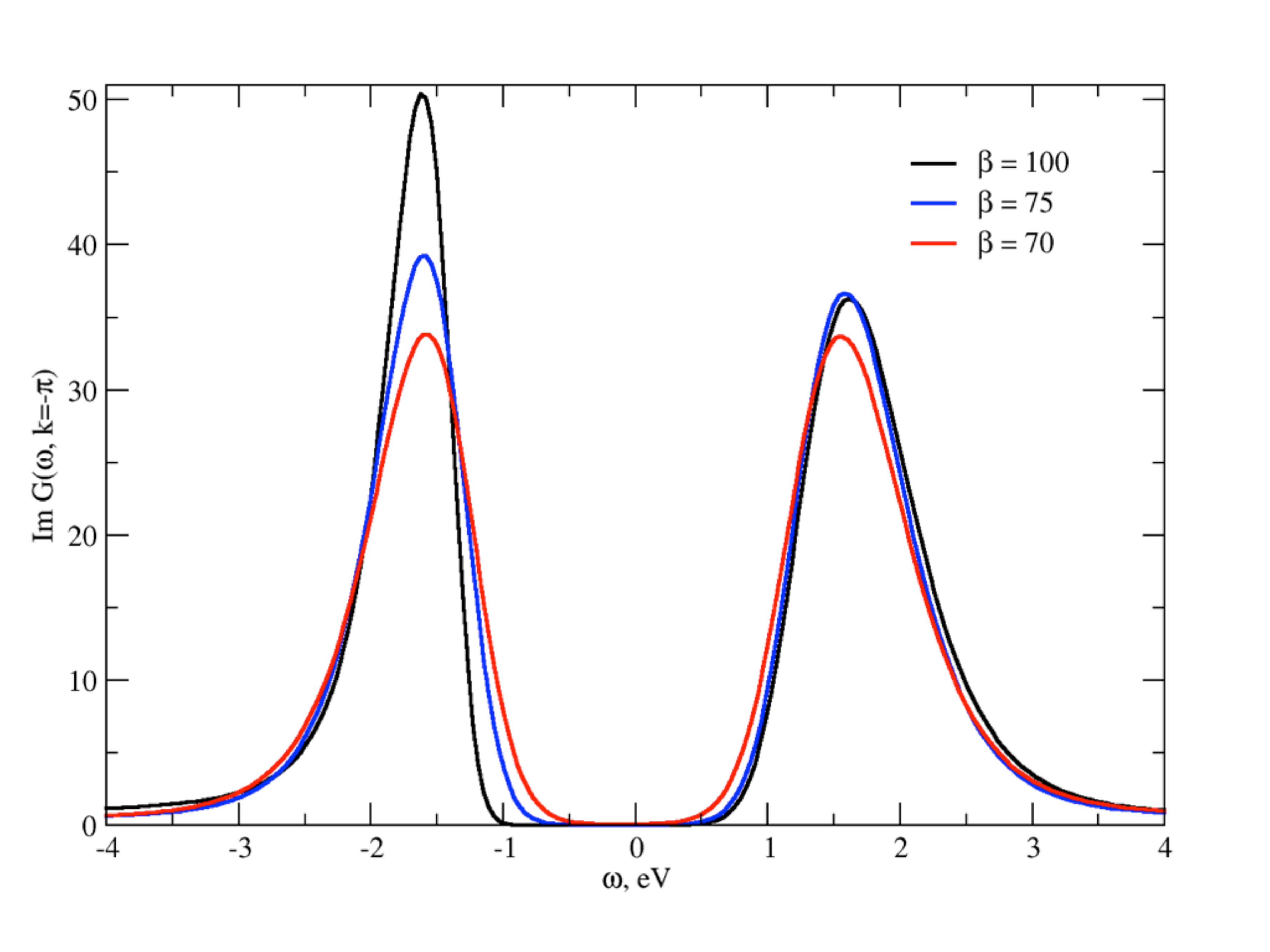}
\caption{Periodic 1D boron nitride solid spectrum for the ``correlated HOCO and LUCO'' (see Sec.~V.~B) for several temperatures at k=$-\pi$. The chemical potential was adjusted for $\omega=0$ to fall in the middle of the band gap.
}
\label{fig:BN_3}
\end{figure}

Shown in Fig.~\ref{fig:BN_proj} is the spectral projection of the real-frequency correlated Green's function at $\beta$=100. Using the analogy previously discussed at the end of Section V.~B, we plot the ``highest occupied'' and ``lowest unoccupied correlated bands'' of 1D BN separated by a sizable gap of approximately 3.5~eV at k=-$\pi$. This magnitude of the band gap is expected based on the properties of its 2D counterpart~\cite{C4CS00102H}. 


Once again, we treat such a simple system as a benchmark example and we are interested in the evolution of the 1D BN band gap as the temperature changes and influences electronic degrees of freedom. Let us stress that while  an evolution of the spectrum as a function of temperature is expected, in order to reproduce it reliably, the computational procedure must be very robust. 
We not only must be able to iteratively converge Green's function and self-energy yielding different Green's functions at different temperatures, we also must continue the results obtained on the imaginary axis to the real axis risking that the continuation will obscure the spectral features and the gap temperature dependence will no longer be visible.

In Fig.~\ref{fig:BN_3} we plot  the spectrum at k=-$\pi$ for $\beta$=100, 75, and 70. As expected, with the increase of temperature (decrease of $\beta$) the spectral peaks corresponding to the bands broaden and the band gap reduces. Note that both Fig.~\ref{fig:BN_proj} and~\ref{fig:BN_3} use eV as energy units for clarity.

\section{Conclusion}\label{sec_conclusions}

While the theory describing the connection between the Matsubara Green's function formalism and thermodynamics has been known since the 1960s, few computational methods are currently capable of employing the Matsubara formalism for realistic systems. This is due to its computationally demanding nature that requires complicated time and frequency grids as well as the iterative nature of the equations. 
Moreover, little is know about obtaining different solutions corresponding to different phases or non-physical solutions that can appear as the result of the non-linear procedure used when Green's functions are constructed iteratively on the imaginary axis.

In our last work~\cite{rusakov}, we have demonstrated one of the first applications of the fully self-consistent Matsubara Green's function formalism to a benchmark periodic problem with realistic interactions, that is, a 1D hydrogen solid.
In the current work, we have shown that it is possible to evaluate temperature dependent thermodynamic quantities using a self-consistent second-order Green's function method. Using the self-consistent Green's function and self-energy, we were able to evaluate the Luttinger-Ward functional at various temperatures and obtain static thermodynamic quantities such as Helmholtz energy, internal energy, and entropy. Evaluation of these quantities gives us access to the partition function and any thermodynamic quantity that can be derived from it. 
Unlike the FCI case, in GF2 we do not need to explicitly calculate excited state energies or Boltzmann factors, making calculation of thermodynamic quantities for larger systems feasible.

To calibrate the thermodynamic data obtained from GF2 against FCI, we have illustrated that for a hydrogen fluoride molecule at high temperature, we are able to obtain energies in excellent agreement with finite temperature FCI. In the lower temperature regime, we observed, as expected,  some deviation from the FCI answer but the overall quality of the results still remained high.

Finally, since the thermodynamic data can be used to construct phase diagrams, we used a simple 1D hydrogen solid to investigate possible phases at different temperatures and interatomic distances.  We obtained different phases such as a band insulator, Mott insulator, or metal, and we were able to distinguish which phase is more stable at various temperatures. Determining the stability is possible due to our ability to calculate not only the internal energy but also the entropic contribution at each temperature. Consequently, based on the difference of the Helmholtz energy, we are able to determine the most stable phase for each of the temperature points. 

Additionally, we have performed a calculation of 1D periodic BN solid demonstrating that GF2 can reproduce the gap deviation as a function of temperature. Thus, for higher temperatures, we have observed the narrowing of the electronic bandgap.

While we acknowledge that GF2 is a low order perturbative method and as such can deliver results that are inaccurate, we believe that for most weakly and moderately correlated systems such as semiconductors with small band gaps, it can be used in the future to provide accurate thermodynamic data and phase diagrams. For systems where the correlations are strong, GF2 can be combined with methods such as self-energy embedding theory (SEET)~\cite{AlexeiSEET,LAN, nguyen2016rigorous} to provide accurate answers concerning thermodynamics. 

\acknowledgments

D.Z. and A.R.W. would like to acknowledge a National Science Foundation (NSF) grant No. CHE-1453894. A.A.R. acknowledges a Department of Energy (DOE) grant No. ER16391 and Prof. Emanuel Gull for helpful discussions.

\section{Appendix: High-frequency expansion of the Green's function for evaluating the Luttinger-Ward functional} \label{Appendix}
Here, we consider the contribution to the Luttinger-Ward functional in the high frequency limit. In the Matsubara formalism for large frequencies, the Green's function and self-energy can be expressed as a series
\begin{equation}
G(i\omega_n) = \frac{G_1}{i\omega_n} + \frac{G_2}{(i\omega_n)^2}+\mathcal{O}(\frac{1}{(i\omega_n)^3}),
\end{equation}
\begin{equation}
\Sigma(i\omega_n) = \frac{\Sigma_1}{i\omega_n} + \frac{\Sigma_2}{(i\omega_n)^2}+\mathcal{O}(\frac{1}{(i\omega_n)^3}).
\end{equation}

\noindent The coefficients for the Green's function expansion for non-orthogonal orbitals with a quantum chemistry Hamiltonian~\cite{rusakov2014local} are given as 

\begin{equation}
\begin{split}
G_1 &= S^{-1}, \\
G_2&= S^{-1}(F-\mu S)S^{-1}.
\end{split}
\end{equation}
The $\Sigma_1$ and $\Sigma_2$ coefficient of the self-energy has a complicated explicit form as demonstrated in Ref.~\onlinecite{rusakov2014local} and is evaluated numerically as

\begin{equation}
\begin{split}
\Sigma_1 &= {\rm Re}\left(\Sigma(i\omega_{max})\times i\omega_{max}\right) \\
\Sigma_2 &=  \left(\Sigma(i\omega_{max})-\frac{\Sigma_1}{i\omega_{max}}\right)\times (i\omega_{max})^2 .
\end{split}
\end{equation}

For the molecular example used in this work, the high frequency contribution to ${\rm Tr}[G \Sigma]$ was evaluated in the same manner as previously discussed in Ref.~\onlinecite{GF2,spline}. At the very high temperatures considered for molecular examples, the high frequency contribution becomes negligible. However, it is necessary to include at the lower temperatures considered. 
The high-frequency contribution to the $\Omega^k_{Tr[G \Sigma]}$ term of Eq.~\ref{eqn:Trace} can be evaluated in a way analogous to the molecular case.

To evaluate the high frequency contribution to the logarithm term ${\rm Tr}[{\rm ln}\{ 1 - G\Sigma \}]$ in Eq.~\ref{eqn:LW_reform}, we expanded the logarithm as a Taylor series ${\rm ln}(1-x) = x - \frac{x^2}{2} + \dots$ to yield

\begin{equation}
{\rm ln}\Big(1-\frac{G_1 \Sigma_1}{(i\omega_n)^2}\Big) = \frac{G_{1}\Sigma_1}{(i\omega_n)^2} + \mathcal{O}\Big(\frac{1}{(i\omega_n)^3} \Big).
\end{equation}
\noindent We exclude in this expansion all terms that are equal or smaller in magnitude than $\mathcal{O}\Big(\frac{1}{(i\omega_n)^3} \Big)$. Thus, it is only necessary practically to evaluate the first term in the expansion to capture the important contribution from the high frequency limit.
We would like to emphasize that although we do not present results including  the high frequency contribution to the logarithm term ${\rm Tr}[{\rm ln}\{ 1 - G\Sigma \}]$ in Eq.~\ref{eqn:LW_reform} for the periodic hydrogen solid, this term was evaluated and was found to be minuscule compared to the magnitude of the other energies. However, we would like stress that it is possible that this contribution may be substantial for other systems.

\section{Supplemental Information}
\begin{table*}
\setlength{\tabcolsep}{5pt}
\renewcommand{\arraystretch}{1.5}
\begin{ruledtabular}

\begin{tabular}{c | c  c c| c c c | c c c  }
& \multicolumn{3}{c |}{  Internal Energy (a.u.) }& \multicolumn{3}{c | }{Grand Potential (a.u.)}& \multicolumn{3 }{c | }{Entropy ($k_B$)} \\
\hline
\underline{T (K)} & \underline{Hartree-Fock}& \underline{GF2 } & \underline{FCI } &\underline{Hartree-Fock} & \underline{ GF2 }   &\underline{ FCI } &
\underline{Hartree-Fock} &\underline{ GF2 } &  \underline{ FCI} \\
$10^{3}$ &-98.571 &-98.588&-98.597 &-98.571 & -98.585 &-99.845&0.0& -0.003 & 0.0 \\
$10^{4}$ & -98.571 & -98.588 & -98.597 & -98.571 & -98.621 & -99.944 & 0.0 & 0.0 & 0.0 \\
$10^{5}$ & -97.944 & -98.135 & -98.049 & -101.021 & -103.067 & -102.107 & 3.175 & 3.566 & 3.475 \\
$10^{6}$ & -96.794& -96.988 & -96.945 & -150.563 & -151.410 & -151.244 & 4.979 & 4.949 & 4.958 \\ 
$10^{7}$ & -92.028 & -92.057 & -92.056 & -729.937 & -730.100 & -730.095 & 5.348 & 5.348 & 5.348 \\ 
$10^{8}$ & -88.483 & -88.487 & -88.487 & -6846.975 & -6847.001 & -6847.003 & 5.406 & 5.406 & 5.406 \\

\end{tabular}
\end{ruledtabular}
\caption{FCI, GF2, and finite temperature Hartree-Fock results for the hydrogen fluoride molecule in STO-3G basis. A higher value of temperature corresponds to a smaller value of $\beta = 1/k_B T$. Note that the  grand potential depends on chemical potential $\Omega=E-TS-\mu N$ that for gaped system is not uniquely defined. The difference in the FCI, finite temperature Hartree-Fock, and GF2 for the $10^4 - 10^4$ K temperatures are the result of this non-uniqueness of the chemical potential. The Helmholtz energy $A=\Omega+\mu N=E-TS$, which is the sum of the grand potential and $\mu N$ term does not suffer from this non-uniqueness. }
\label{tab:FH_molecule}
\end{table*}
\clearpage


\begin{thebibliography}{58}%
\makeatletter
\providecommand \@ifxundefined [1]{%
 \@ifx{#1\undefined}
}%
\providecommand \@ifnum [1]{%
 \ifnum #1\expandafter \@firstoftwo
 \else \expandafter \@secondoftwo
 \fi
}%
\providecommand \@ifx [1]{%
 \ifx #1\expandafter \@firstoftwo
 \else \expandafter \@secondoftwo
 \fi
}%
\providecommand \natexlab [1]{#1}%
\providecommand \enquote  [1]{``#1''}%
\providecommand \bibnamefont  [1]{#1}%
\providecommand \bibfnamefont [1]{#1}%
\providecommand \citenamefont [1]{#1}%
\providecommand \href@noop [0]{\@secondoftwo}%
\providecommand \href [0]{\begingroup \@sanitize@url \@href}%
\providecommand \@href[1]{\@@startlink{#1}\@@href}%
\providecommand \@@href[1]{\endgroup#1\@@endlink}%
\providecommand \@sanitize@url [0]{\catcode `\\12\catcode `\$12\catcode
  `\&12\catcode `\#12\catcode `\^12\catcode `\_12\catcode `\%12\relax}%
\providecommand \@@startlink[1]{}%
\providecommand \@@endlink[0]{}%
\providecommand \url  [0]{\begingroup\@sanitize@url \@url }%
\providecommand \@url [1]{\endgroup\@href {#1}{\urlprefix }}%
\providecommand \urlprefix  [0]{URL }%
\providecommand \Eprint [0]{\href }%
\providecommand \doibase [0]{http://dx.doi.org/}%
\providecommand \selectlanguage [0]{\@gobble}%
\providecommand \bibinfo  [0]{\@secondoftwo}%
\providecommand \bibfield  [0]{\@secondoftwo}%
\providecommand \translation [1]{[#1]}%
\providecommand \BibitemOpen [0]{}%
\providecommand \bibitemStop [0]{}%
\providecommand \bibitemNoStop [0]{.\EOS\space}%
\providecommand \EOS [0]{\spacefactor3000\relax}%
\providecommand \BibitemShut  [1]{\csname bibitem#1\endcsname}%
\let\auto@bib@innerbib\@empty
\bibitem [{\citenamefont {McQuarrie}\ and\ \citenamefont
  {Simon}(1999)}]{mcquarrie1999molecular}%
  \BibitemOpen
  \bibfield  {author} {\bibinfo {author} {\bibfnamefont {D.~A.}\ \bibnamefont
  {McQuarrie}}\ and\ \bibinfo {author} {\bibfnamefont {J.~D.}\ \bibnamefont
  {Simon}},\ }\href@noop {} {\emph {\bibinfo {title} {Molecular
  thermodynamics}}}\ (\bibinfo  {publisher} {University Science Books
  Sausalito, CA},\ \bibinfo {year} {1999})\BibitemShut {NoStop}%
\bibitem [{\citenamefont {Ochterski}(2000)}]{g09thermo}%
  \BibitemOpen
  \bibfield  {author} {\bibinfo {author} {\bibfnamefont {J.~W.}\ \bibnamefont
  {Ochterski}},\ }\href@noop {} {\bibfield  {journal} {\bibinfo  {journal}
  {Gaussian Inc, Pittsburgh, PA}\ ,\ \bibinfo {pages} {1}} (\bibinfo {year}
  {2000})}\BibitemShut {NoStop}%
\bibitem [{\citenamefont {Tsao}()}]{guidebookworld}%
  \BibitemOpen
  \bibfield  {author} {\bibinfo {author} {\bibfnamefont {J.~Y.}\ \bibnamefont
  {Tsao}},\ }\href@noop {} {\emph {\bibinfo {title} {The World of Compound
  Semiconductors}}}\ (\bibinfo  {publisher} {Citeseer})\BibitemShut {NoStop}%
\bibitem [{\citenamefont {Moroni}\ \emph {et~al.}(1996)\citenamefont {Moroni},
  \citenamefont {Grimvall},\ and\ \citenamefont {Jarlborg}}]{metalsfreeenergy}%
  \BibitemOpen
  \bibfield  {author} {\bibinfo {author} {\bibfnamefont {E.~G.}\ \bibnamefont
  {Moroni}}, \bibinfo {author} {\bibfnamefont {G.}~\bibnamefont {Grimvall}}, \
  and\ \bibinfo {author} {\bibfnamefont {T.}~\bibnamefont {Jarlborg}},\
  }\href@noop {} {\bibfield  {journal} {\bibinfo  {journal} {Physical review
  letters}\ }\textbf {\bibinfo {volume} {76}},\ \bibinfo {pages} {2758}
  (\bibinfo {year} {1996})}\BibitemShut {NoStop}%
\bibitem [{\citenamefont {Yu}\ and\ \citenamefont
  {Tuckerman}(2011)}]{benzenetuckerman}%
  \BibitemOpen
  \bibfield  {author} {\bibinfo {author} {\bibfnamefont {T.-Q.}\ \bibnamefont
  {Yu}}\ and\ \bibinfo {author} {\bibfnamefont {M.~E.}\ \bibnamefont
  {Tuckerman}},\ }\href@noop {} {\bibfield  {journal} {\bibinfo  {journal}
  {Physical review letters}\ }\textbf {\bibinfo {volume} {107}},\ \bibinfo
  {pages} {015701} (\bibinfo {year} {2011})}\BibitemShut {NoStop}%
\bibitem [{\citenamefont {Zhu}\ \emph {et~al.}(2016)\citenamefont {Zhu},
  \citenamefont {Shtukenberg}, \citenamefont {Carter}, \citenamefont {Yu},
  \citenamefont {Yang}, \citenamefont {Chen}, \citenamefont {Raiteri},
  \citenamefont {Oganov}, \citenamefont {Pokroy}, \citenamefont {Polishchuk}
  \emph {et~al.}}]{resorcinol}%
  \BibitemOpen
  \bibfield  {author} {\bibinfo {author} {\bibfnamefont {Q.}~\bibnamefont
  {Zhu}}, \bibinfo {author} {\bibfnamefont {A.~G.}\ \bibnamefont
  {Shtukenberg}}, \bibinfo {author} {\bibfnamefont {D.~J.}\ \bibnamefont
  {Carter}}, \bibinfo {author} {\bibfnamefont {T.-Q.}\ \bibnamefont {Yu}},
  \bibinfo {author} {\bibfnamefont {J.}~\bibnamefont {Yang}}, \bibinfo {author}
  {\bibfnamefont {M.}~\bibnamefont {Chen}}, \bibinfo {author} {\bibfnamefont
  {P.}~\bibnamefont {Raiteri}}, \bibinfo {author} {\bibfnamefont {A.~R.}\
  \bibnamefont {Oganov}}, \bibinfo {author} {\bibfnamefont {B.}~\bibnamefont
  {Pokroy}}, \bibinfo {author} {\bibfnamefont {I.}~\bibnamefont {Polishchuk}},
  \emph {et~al.},\ }\href@noop {} {\bibfield  {journal} {\bibinfo  {journal}
  {Journal of the American Chemical Society}\ }\textbf {\bibinfo {volume}
  {138}},\ \bibinfo {pages} {4881} (\bibinfo {year} {2016})}\BibitemShut
  {NoStop}%
\bibitem [{\citenamefont {Mermin}(1963)}]{mermin1963stability}%
  \BibitemOpen
  \bibfield  {author} {\bibinfo {author} {\bibfnamefont {N.~D.}\ \bibnamefont
  {Mermin}},\ }\href@noop {} {\bibfield  {journal} {\bibinfo  {journal} {Annals
  of Physics}\ }\textbf {\bibinfo {volume} {21}},\ \bibinfo {pages} {99}
  (\bibinfo {year} {1963})}\BibitemShut {NoStop}%
\bibitem [{\citenamefont {Mermin}(1965)}]{mermin1965thermal}%
  \BibitemOpen
  \bibfield  {author} {\bibinfo {author} {\bibfnamefont {N.~D.}\ \bibnamefont
  {Mermin}},\ }\href@noop {} {\bibfield  {journal} {\bibinfo  {journal}
  {Physical Review}\ }\textbf {\bibinfo {volume} {137}},\ \bibinfo {pages}
  {A1441} (\bibinfo {year} {1965})}\BibitemShut {NoStop}%
\bibitem [{\citenamefont {Dharma-wardana}(2016)}]{dharma2016current}%
  \BibitemOpen
  \bibfield  {author} {\bibinfo {author} {\bibfnamefont {M.}~\bibnamefont
  {Dharma-wardana}},\ }\href@noop {} {\bibfield  {journal} {\bibinfo  {journal}
  {arXiv preprint arXiv:1602.04734}\ } (\bibinfo {year} {2016})}\BibitemShut
  {NoStop}%
\bibitem [{\citenamefont {Smith}\ \emph {et~al.}(2015)\citenamefont {Smith},
  \citenamefont {Pribram-Jones},\ and\ \citenamefont
  {Burke}}]{smith2015thermal}%
  \BibitemOpen
  \bibfield  {author} {\bibinfo {author} {\bibfnamefont {J.}~\bibnamefont
  {Smith}}, \bibinfo {author} {\bibfnamefont {A.}~\bibnamefont
  {Pribram-Jones}}, \ and\ \bibinfo {author} {\bibfnamefont {K.}~\bibnamefont
  {Burke}},\ }\href@noop {} {\bibfield  {journal} {\bibinfo  {journal} {arXiv
  preprint arXiv:1509.03097}\ } (\bibinfo {year} {2015})}\BibitemShut {NoStop}%
\bibitem [{\citenamefont {Karasiev}\ \emph {et~al.}(2014)\citenamefont
  {Karasiev}, \citenamefont {Sjostrom}, \citenamefont {Dufty},\ and\
  \citenamefont {Trickey}}]{trickey}%
  \BibitemOpen
  \bibfield  {author} {\bibinfo {author} {\bibfnamefont {V.~V.}\ \bibnamefont
  {Karasiev}}, \bibinfo {author} {\bibfnamefont {T.}~\bibnamefont {Sjostrom}},
  \bibinfo {author} {\bibfnamefont {J.}~\bibnamefont {Dufty}}, \ and\ \bibinfo
  {author} {\bibfnamefont {S.}~\bibnamefont {Trickey}},\ }\href@noop {}
  {\bibfield  {journal} {\bibinfo  {journal} {Physical review letters}\
  }\textbf {\bibinfo {volume} {112}},\ \bibinfo {pages} {076403} (\bibinfo
  {year} {2014})}\BibitemShut {NoStop}%
\bibitem [{\citenamefont {Hirata}\ and\ \citenamefont
  {He}(2013)}]{hirata2013kohn}%
  \BibitemOpen
  \bibfield  {author} {\bibinfo {author} {\bibfnamefont {S.}~\bibnamefont
  {Hirata}}\ and\ \bibinfo {author} {\bibfnamefont {X.}~\bibnamefont {He}},\
  }\href@noop {} {\bibfield  {journal} {\bibinfo  {journal} {The Journal of
  chemical physics}\ }\textbf {\bibinfo {volume} {138}},\ \bibinfo {pages}
  {204112} (\bibinfo {year} {2013})}\BibitemShut {NoStop}%
\bibitem [{\citenamefont {Altenbokum}\ \emph {et~al.}(1987)\citenamefont
  {Altenbokum}, \citenamefont {Emrich}, \citenamefont {K{\"u}mmel},\ and\
  \citenamefont {Zabolitzky}}]{altenbokum1987temperature}%
  \BibitemOpen
  \bibfield  {author} {\bibinfo {author} {\bibfnamefont {M.}~\bibnamefont
  {Altenbokum}}, \bibinfo {author} {\bibfnamefont {K.}~\bibnamefont {Emrich}},
  \bibinfo {author} {\bibfnamefont {H.}~\bibnamefont {K{\"u}mmel}}, \ and\
  \bibinfo {author} {\bibfnamefont {J.}~\bibnamefont {Zabolitzky}},\ }in\
  \href@noop {} {\emph {\bibinfo {booktitle} {Condensed matter theories}}}\
  (\bibinfo  {publisher} {Springer},\ \bibinfo {year} {1987})\ pp.\ \bibinfo
  {pages} {389--396}\BibitemShut {NoStop}%
\bibitem [{\citenamefont {Mandal}\ \emph {et~al.}(2003)\citenamefont {Mandal},
  \citenamefont {Ghosh}, \citenamefont {Sanyal},\ and\ \citenamefont
  {Mukherjee}}]{mandal2003finite}%
  \BibitemOpen
  \bibfield  {author} {\bibinfo {author} {\bibfnamefont {S.~H.}\ \bibnamefont
  {Mandal}}, \bibinfo {author} {\bibfnamefont {R.}~\bibnamefont {Ghosh}},
  \bibinfo {author} {\bibfnamefont {G.}~\bibnamefont {Sanyal}}, \ and\ \bibinfo
  {author} {\bibfnamefont {D.}~\bibnamefont {Mukherjee}},\ }\href@noop {}
  {\bibfield  {journal} {\bibinfo  {journal} {International Journal of Modern
  Physics B}\ }\textbf {\bibinfo {volume} {17}},\ \bibinfo {pages} {5367}
  (\bibinfo {year} {2003})}\BibitemShut {NoStop}%
\bibitem [{\citenamefont {Hermes}\ and\ \citenamefont
  {Hirata}(2015)}]{sohiratacc}%
  \BibitemOpen
  \bibfield  {author} {\bibinfo {author} {\bibfnamefont {M.~R.}\ \bibnamefont
  {Hermes}}\ and\ \bibinfo {author} {\bibfnamefont {S.}~\bibnamefont
  {Hirata}},\ }\href@noop {} {\bibfield  {journal} {\bibinfo  {journal} {The
  Journal of Chemical Physics}\ }\textbf {\bibinfo {volume} {143}},\ \bibinfo
  {pages} {102818} (\bibinfo {year} {2015})}\BibitemShut {NoStop}%
\bibitem [{\citenamefont {Jakli{\v{c}}}\ and\ \citenamefont
  {Prelov{\v{s}}ek}(1994)}]{jaklivc1994lanczos}%
  \BibitemOpen
  \bibfield  {author} {\bibinfo {author} {\bibfnamefont {J.}~\bibnamefont
  {Jakli{\v{c}}}}\ and\ \bibinfo {author} {\bibfnamefont {P.}~\bibnamefont
  {Prelov{\v{s}}ek}},\ }\href@noop {} {\bibfield  {journal} {\bibinfo
  {journal} {Physical Review B}\ }\textbf {\bibinfo {volume} {49}},\ \bibinfo
  {pages} {5065} (\bibinfo {year} {1994})}\BibitemShut {NoStop}%
\bibitem [{\citenamefont {Prelovsek}(2011)}]{ftlanczos}%
  \BibitemOpen
  \bibfield  {author} {\bibinfo {author} {\bibfnamefont {J.}~\bibnamefont
  {Prelovsek}, \bibfnamefont {P.~Bonca}},\ }\href@noop {} {\bibfield  {journal}
  {\bibinfo  {journal} {arXiv preprint arXiv:1111.5931}\ } (\bibinfo {year}
  {2011})}\BibitemShut {NoStop}%
\bibitem [{\citenamefont {Stefanucci}\ and\ \citenamefont {van
  Leeuwen}(2013)}]{vanleeuwentext}%
  \BibitemOpen
  \bibfield  {author} {\bibinfo {author} {\bibfnamefont {G.}~\bibnamefont
  {Stefanucci}}\ and\ \bibinfo {author} {\bibfnamefont {R.}~\bibnamefont {van
  Leeuwen}},\ }\href@noop {} {\emph {\bibinfo {title} {Nonequilibrium Many-Body
  Theory of Quantum Systems: A Modern Introduction}}}\ (\bibinfo  {publisher}
  {Cambridge University Press},\ \bibinfo {year} {2013})\BibitemShut {NoStop}%
\bibitem [{\citenamefont {Mattuck}(2012)}]{mattuck2012guide}%
  \BibitemOpen
  \bibfield  {author} {\bibinfo {author} {\bibfnamefont {R.~D.}\ \bibnamefont
  {Mattuck}},\ }\href@noop {} {\emph {\bibinfo {title} {A Guide to Feynman
  Diagrams in the Many-Body Problem}}}\ (\bibinfo  {publisher} {Courier
  Corporation},\ \bibinfo {year} {2012})\BibitemShut {NoStop}%
\bibitem [{\citenamefont {Fetter}\ and\ \citenamefont
  {Walecka}(2003)}]{fetter2003quantum}%
  \BibitemOpen
  \bibfield  {author} {\bibinfo {author} {\bibfnamefont {A.~L.}\ \bibnamefont
  {Fetter}}\ and\ \bibinfo {author} {\bibfnamefont {J.~D.}\ \bibnamefont
  {Walecka}},\ }\href@noop {} {\emph {\bibinfo {title} {Quantum theory of
  many-particle systems}}}\ (\bibinfo  {publisher} {Courier Corporation},\
  \bibinfo {year} {2003})\BibitemShut {NoStop}%
\bibitem [{\citenamefont {Dzyaloshinski}\ \emph {et~al.}(1975)\citenamefont
  {Dzyaloshinski} \emph {et~al.}}]{dzyaloshinski1975methods}%
  \BibitemOpen
  \bibfield  {author} {\bibinfo {author} {\bibfnamefont {I.}~\bibnamefont
  {Dzyaloshinski}} \emph {et~al.},\ }\href@noop {} {\emph {\bibinfo {title}
  {Methods of quantum field theory in statistical physics}}}\ (\bibinfo
  {publisher} {Courier Corporation},\ \bibinfo {year} {1975})\BibitemShut
  {NoStop}%
\bibitem [{\citenamefont {Li}\ \emph {et~al.}(2009)\citenamefont {Li},
  \citenamefont {Hanke}, \citenamefont {Rubtsov}, \citenamefont {B{\"a}se},\
  and\ \citenamefont {Potthoff}}]{potthoff1}%
  \BibitemOpen
  \bibfield  {author} {\bibinfo {author} {\bibfnamefont {G.}~\bibnamefont
  {Li}}, \bibinfo {author} {\bibfnamefont {W.}~\bibnamefont {Hanke}}, \bibinfo
  {author} {\bibfnamefont {A.~N.}\ \bibnamefont {Rubtsov}}, \bibinfo {author}
  {\bibfnamefont {S.}~\bibnamefont {B{\"a}se}}, \ and\ \bibinfo {author}
  {\bibfnamefont {M.}~\bibnamefont {Potthoff}},\ }\href@noop {} {\bibfield
  {journal} {\bibinfo  {journal} {Physical Review B}\ }\textbf {\bibinfo
  {volume} {80}},\ \bibinfo {pages} {195118} (\bibinfo {year}
  {2009})}\BibitemShut {NoStop}%
\bibitem [{\citenamefont {Potthoff}\ \emph {et~al.}(2003)\citenamefont
  {Potthoff}, \citenamefont {Aichhorn},\ and\ \citenamefont
  {Dahnken}}]{potthoff2}%
  \BibitemOpen
  \bibfield  {author} {\bibinfo {author} {\bibfnamefont {M.}~\bibnamefont
  {Potthoff}}, \bibinfo {author} {\bibfnamefont {M.}~\bibnamefont {Aichhorn}},
  \ and\ \bibinfo {author} {\bibfnamefont {C.}~\bibnamefont {Dahnken}},\
  }\href@noop {} {\bibfield  {journal} {\bibinfo  {journal} {Physical Review
  Letters}\ }\textbf {\bibinfo {volume} {91}},\ \bibinfo {pages} {206402}
  (\bibinfo {year} {2003})}\BibitemShut {NoStop}%
\bibitem [{\citenamefont {Potthoff}(2003{\natexlab{a}})}]{potthoff2003Hubbard}%
  \BibitemOpen
  \bibfield  {author} {\bibinfo {author} {\bibfnamefont {M.}~\bibnamefont
  {Potthoff}},\ }\href@noop {} {\bibfield  {journal} {\bibinfo  {journal} {The
  European Physical Journal B-Condensed Matter and Complex Systems}\ }\textbf
  {\bibinfo {volume} {36}},\ \bibinfo {pages} {335} (\bibinfo {year}
  {2003}{\natexlab{a}})}\BibitemShut {NoStop}%
\bibitem [{\citenamefont {Potthoff}()}]{PotthoffSE}%
  \BibitemOpen
  \bibfield  {author} {\bibinfo {author} {\bibfnamefont {M.}~\bibnamefont
  {Potthoff}},\ }\href@noop {} {\bibinfo  {journal} {arXiv:1407.4065
  [cond-mat.str-el]}\ }\BibitemShut {NoStop}%
\bibitem [{\citenamefont {Potthoff}(2003{\natexlab{b}})}]{potthoff2003self}%
  \BibitemOpen
\bibfield  {journal} {  }\bibfield  {author} {\bibinfo {author} {\bibfnamefont
  {M.}~\bibnamefont {Potthoff}},\ }\href@noop {} {\bibfield  {journal}
  {\bibinfo  {journal} {The European Physical Journal B-Condensed Matter and
  Complex Systems}\ }\textbf {\bibinfo {volume} {32}},\ \bibinfo {pages} {429}
  (\bibinfo {year} {2003}{\natexlab{b}})}\BibitemShut {NoStop}%
\bibitem [{\citenamefont {Neuhauser}\ \emph {et~al.}(2014)\citenamefont
  {Neuhauser}, \citenamefont {Gao}, \citenamefont {Arntsen}, \citenamefont
  {Karshenas}, \citenamefont {Rabani},\ and\ \citenamefont
  {Baer}}]{neuhauser2014breaking}%
  \BibitemOpen
  \bibfield  {author} {\bibinfo {author} {\bibfnamefont {D.}~\bibnamefont
  {Neuhauser}}, \bibinfo {author} {\bibfnamefont {Y.}~\bibnamefont {Gao}},
  \bibinfo {author} {\bibfnamefont {C.}~\bibnamefont {Arntsen}}, \bibinfo
  {author} {\bibfnamefont {C.}~\bibnamefont {Karshenas}}, \bibinfo {author}
  {\bibfnamefont {E.}~\bibnamefont {Rabani}}, \ and\ \bibinfo {author}
  {\bibfnamefont {R.}~\bibnamefont {Baer}},\ }\href@noop {} {\bibfield
  {journal} {\bibinfo  {journal} {Physical review letters}\ }\textbf {\bibinfo
  {volume} {113}},\ \bibinfo {pages} {076402} (\bibinfo {year}
  {2014})}\BibitemShut {NoStop}%
\bibitem [{\citenamefont {Hung}\ \emph {et~al.}(2016)\citenamefont {Hung},
  \citenamefont {Felipe}, \citenamefont {Souto-Casares}, \citenamefont
  {Chelikowsky}, \citenamefont {Louie},\ and\ \citenamefont
  {{\"O}{\u{g}}{\"u}t}}]{hung2016excitation}%
  \BibitemOpen
  \bibfield  {author} {\bibinfo {author} {\bibfnamefont {L.}~\bibnamefont
  {Hung}}, \bibinfo {author} {\bibfnamefont {H.}~\bibnamefont {Felipe}},
  \bibinfo {author} {\bibfnamefont {J.}~\bibnamefont {Souto-Casares}}, \bibinfo
  {author} {\bibfnamefont {J.~R.}\ \bibnamefont {Chelikowsky}}, \bibinfo
  {author} {\bibfnamefont {S.~G.}\ \bibnamefont {Louie}}, \ and\ \bibinfo
  {author} {\bibfnamefont {S.}~\bibnamefont {{\"O}{\u{g}}{\"u}t}},\ }\href@noop
  {} {\bibfield  {journal} {\bibinfo  {journal} {Physical Review B}\ }\textbf
  {\bibinfo {volume} {94}},\ \bibinfo {pages} {085125} (\bibinfo {year}
  {2016})}\BibitemShut {NoStop}%
\bibitem [{\citenamefont {van Schilfgaarde}\ \emph {et~al.}(2006)\citenamefont
  {van Schilfgaarde}, \citenamefont {Kotani},\ and\ \citenamefont
  {Faleev}}]{van2006quasiparticle}%
  \BibitemOpen
  \bibfield  {author} {\bibinfo {author} {\bibfnamefont {M.}~\bibnamefont {van
  Schilfgaarde}}, \bibinfo {author} {\bibfnamefont {T.}~\bibnamefont {Kotani}},
  \ and\ \bibinfo {author} {\bibfnamefont {S.}~\bibnamefont {Faleev}},\
  }\href@noop {} {\bibfield  {journal} {\bibinfo  {journal} {Physical review
  letters}\ }\textbf {\bibinfo {volume} {96}},\ \bibinfo {pages} {226402}
  (\bibinfo {year} {2006})}\BibitemShut {NoStop}%
\bibitem [{\citenamefont {Caruso}\ \emph {et~al.}(2013)\citenamefont {Caruso},
  \citenamefont {Rinke}, \citenamefont {Ren}, \citenamefont {Rubio},\ and\
  \citenamefont {Scheffler}}]{caruso2013self}%
  \BibitemOpen
  \bibfield  {author} {\bibinfo {author} {\bibfnamefont {F.}~\bibnamefont
  {Caruso}}, \bibinfo {author} {\bibfnamefont {P.}~\bibnamefont {Rinke}},
  \bibinfo {author} {\bibfnamefont {X.}~\bibnamefont {Ren}}, \bibinfo {author}
  {\bibfnamefont {A.}~\bibnamefont {Rubio}}, \ and\ \bibinfo {author}
  {\bibfnamefont {M.}~\bibnamefont {Scheffler}},\ }\href@noop {} {\bibfield
  {journal} {\bibinfo  {journal} {Physical Review B}\ }\textbf {\bibinfo
  {volume} {88}},\ \bibinfo {pages} {075105} (\bibinfo {year}
  {2013})}\BibitemShut {NoStop}%
\bibitem [{\citenamefont {Faleev}\ \emph {et~al.}(2004)\citenamefont {Faleev},
  \citenamefont {Van~Schilfgaarde},\ and\ \citenamefont
  {Kotani}}]{faleev2004all}%
  \BibitemOpen
  \bibfield  {author} {\bibinfo {author} {\bibfnamefont {S.~V.}\ \bibnamefont
  {Faleev}}, \bibinfo {author} {\bibfnamefont {M.}~\bibnamefont
  {Van~Schilfgaarde}}, \ and\ \bibinfo {author} {\bibfnamefont
  {T.}~\bibnamefont {Kotani}},\ }\href@noop {} {\bibfield  {journal} {\bibinfo
  {journal} {Physical review letters}\ }\textbf {\bibinfo {volume} {93}},\
  \bibinfo {pages} {126406} (\bibinfo {year} {2004})}\BibitemShut {NoStop}%
\bibitem [{\citenamefont {Rostgaard}\ \emph {et~al.}(2010)\citenamefont
  {Rostgaard}, \citenamefont {Jacobsen},\ and\ \citenamefont
  {Thygesen}}]{rostgaard2010fully}%
  \BibitemOpen
  \bibfield  {author} {\bibinfo {author} {\bibfnamefont {C.}~\bibnamefont
  {Rostgaard}}, \bibinfo {author} {\bibfnamefont {K.~W.}\ \bibnamefont
  {Jacobsen}}, \ and\ \bibinfo {author} {\bibfnamefont {K.~S.}\ \bibnamefont
  {Thygesen}},\ }\href@noop {} {\bibfield  {journal} {\bibinfo  {journal}
  {Physical Review B}\ }\textbf {\bibinfo {volume} {81}},\ \bibinfo {pages}
  {085103} (\bibinfo {year} {2010})}\BibitemShut {NoStop}%
\bibitem [{\citenamefont {Govoni}\ and\ \citenamefont
  {Galli}(2015)}]{govoni2015large}%
  \BibitemOpen
  \bibfield  {author} {\bibinfo {author} {\bibfnamefont {M.}~\bibnamefont
  {Govoni}}\ and\ \bibinfo {author} {\bibfnamefont {G.}~\bibnamefont {Galli}},\
  }\href@noop {} {\bibfield  {journal} {\bibinfo  {journal} {Journal of
  chemical theory and computation}\ }\textbf {\bibinfo {volume} {11}},\
  \bibinfo {pages} {2680} (\bibinfo {year} {2015})}\BibitemShut {NoStop}%
\bibitem [{\citenamefont {van Leeuwen}\ \emph {et~al.}(2006)\citenamefont {van
  Leeuwen}, \citenamefont {Dahlen},\ and\ \citenamefont {Stan}}]{van2006total}%
  \BibitemOpen
  \bibfield  {author} {\bibinfo {author} {\bibfnamefont {R.}~\bibnamefont {van
  Leeuwen}}, \bibinfo {author} {\bibfnamefont {N.~E.}\ \bibnamefont {Dahlen}},
  \ and\ \bibinfo {author} {\bibfnamefont {A.}~\bibnamefont {Stan}},\
  }\href@noop {} {\bibfield  {journal} {\bibinfo  {journal} {Physical Review
  B}\ }\textbf {\bibinfo {volume} {74}},\ \bibinfo {pages} {195105} (\bibinfo
  {year} {2006})}\BibitemShut {NoStop}%
\bibitem [{\citenamefont {Dahlen}\ \emph {et~al.}(2006)\citenamefont {Dahlen},
  \citenamefont {van Leeuwen},\ and\ \citenamefont {von
  Barth}}]{dahlen2006variational}%
  \BibitemOpen
  \bibfield  {author} {\bibinfo {author} {\bibfnamefont {N.~E.}\ \bibnamefont
  {Dahlen}}, \bibinfo {author} {\bibfnamefont {R.}~\bibnamefont {van Leeuwen}},
  \ and\ \bibinfo {author} {\bibfnamefont {U.}~\bibnamefont {von Barth}},\
  }\href@noop {} {\bibfield  {journal} {\bibinfo  {journal} {Physical Review
  A}\ }\textbf {\bibinfo {volume} {73}},\ \bibinfo {pages} {012511} (\bibinfo
  {year} {2006})}\BibitemShut {NoStop}%
\bibitem [{\citenamefont {Phillips}\ and\ \citenamefont {Zgid}(2014)}]{GF2}%
  \BibitemOpen
  \bibfield  {author} {\bibinfo {author} {\bibfnamefont {J.~J.}\ \bibnamefont
  {Phillips}}\ and\ \bibinfo {author} {\bibfnamefont {D.}~\bibnamefont
  {Zgid}},\ }\href@noop {} {\bibfield  {journal} {\bibinfo  {journal} {The
  Journal of Chemical Physics}\ }\textbf {\bibinfo {volume} {140}},\ \bibinfo
  {pages} {241101} (\bibinfo {year} {2014})}\BibitemShut {NoStop}%
\bibitem [{\citenamefont {Rusakov}\ and\ \citenamefont {Zgid}(2016)}]{rusakov}%
  \BibitemOpen
  \bibfield  {author} {\bibinfo {author} {\bibfnamefont {A.~A.}\ \bibnamefont
  {Rusakov}}\ and\ \bibinfo {author} {\bibfnamefont {D.}~\bibnamefont {Zgid}},\
  }\href@noop {} {\bibfield  {journal} {\bibinfo  {journal} {The Journal of
  chemical physics}\ }\textbf {\bibinfo {volume} {144}},\ \bibinfo {pages}
  {054106} (\bibinfo {year} {2016})}\BibitemShut {NoStop}%
\bibitem [{\citenamefont {Kananenka}\ \emph
  {et~al.}(2016{\natexlab{a}})\citenamefont {Kananenka}, \citenamefont
  {Welden}, \citenamefont {Lan}, \citenamefont {Gull},\ and\ \citenamefont
  {Zgid}}]{spline}%
  \BibitemOpen
  \bibfield  {author} {\bibinfo {author} {\bibfnamefont {A.~A.}\ \bibnamefont
  {Kananenka}}, \bibinfo {author} {\bibfnamefont {A.~R.}\ \bibnamefont
  {Welden}}, \bibinfo {author} {\bibfnamefont {T.~N.}\ \bibnamefont {Lan}},
  \bibinfo {author} {\bibfnamefont {E.}~\bibnamefont {Gull}}, \ and\ \bibinfo
  {author} {\bibfnamefont {D.}~\bibnamefont {Zgid}},\ }\href@noop {} {\bibfield
   {journal} {\bibinfo  {journal} {Journal of Chemical Theory and Computation}\
  } (\bibinfo {year} {2016}{\natexlab{a}})}\BibitemShut {NoStop}%
\bibitem [{\citenamefont {Kananenka}\ \emph {et~al.}(2015)\citenamefont
  {Kananenka}, \citenamefont {Gull},\ and\ \citenamefont {Zgid}}]{AlexeiSEET}%
  \BibitemOpen
  \bibfield  {author} {\bibinfo {author} {\bibfnamefont {A.~A.}\ \bibnamefont
  {Kananenka}}, \bibinfo {author} {\bibfnamefont {E.}~\bibnamefont {Gull}}, \
  and\ \bibinfo {author} {\bibfnamefont {D.}~\bibnamefont {Zgid}},\ }\href@noop
  {} {\bibfield  {journal} {\bibinfo  {journal} {Physical Review B}\ }\textbf
  {\bibinfo {volume} {91}},\ \bibinfo {pages} {121111} (\bibinfo {year}
  {2015})}\BibitemShut {NoStop}%
\bibitem [{\citenamefont {Dickhoff}\ and\ \citenamefont
  {Van~Neck}(2008)}]{vanneck}%
  \BibitemOpen
  \bibfield  {author} {\bibinfo {author} {\bibfnamefont {W.~H.}\ \bibnamefont
  {Dickhoff}}\ and\ \bibinfo {author} {\bibfnamefont {D.}~\bibnamefont
  {Van~Neck}},\ }\href@noop {} {\emph {\bibinfo {title} {Many-body theory
  exposed!: propagator description of quantum mechanics in many-body
  systems}}}\ (\bibinfo  {publisher} {World Scientific},\ \bibinfo {year}
  {2008})\BibitemShut {NoStop}%
\bibitem [{\citenamefont {Luttinger}\ and\ \citenamefont
  {Ward}(1960)}]{LW1960}%
  \BibitemOpen
  \bibfield  {author} {\bibinfo {author} {\bibfnamefont {J.~M.}\ \bibnamefont
  {Luttinger}}\ and\ \bibinfo {author} {\bibfnamefont {J.~C.}\ \bibnamefont
  {Ward}},\ }\href@noop {} {\bibfield  {journal} {\bibinfo  {journal} {Physical
  Review}\ }\textbf {\bibinfo {volume} {118}},\ \bibinfo {pages} {1417}
  (\bibinfo {year} {1960})}\BibitemShut {NoStop}%
\bibitem [{foo({\natexlab{a}})}]{footnote2}%
  \BibitemOpen
  \href@noop {} {} ({\natexlab{a}}),\ \bibinfo {note} {in some texts the
  functional $\Phi$ is simply referred to as the Phi functional and the
  functional $\Omega$ is called the Luttinger-Ward functional.}\BibitemShut
  {Stop}%
\bibitem [{\citenamefont {Almbladh}\ \emph {et~al.}(1999)\citenamefont
  {Almbladh}, \citenamefont {Barth},\ and\ \citenamefont {Leeuwen}}]{VE}%
  \BibitemOpen
  \bibfield  {author} {\bibinfo {author} {\bibfnamefont {C.-O.}\ \bibnamefont
  {Almbladh}}, \bibinfo {author} {\bibfnamefont {U.~V.}\ \bibnamefont {Barth}},
  \ and\ \bibinfo {author} {\bibfnamefont {R.~v.}\ \bibnamefont {Leeuwen}},\
  }\href@noop {} {\bibfield  {journal} {\bibinfo  {journal} {International
  Journal of Modern Physics B}\ }\textbf {\bibinfo {volume} {13}},\ \bibinfo
  {pages} {535} (\bibinfo {year} {1999})}\BibitemShut {NoStop}%
\bibitem [{\citenamefont {Jani\ifmmode~\check{s}\else
  \v{s}\fi{}}(1999)}]{Janis}%
  \BibitemOpen
  \bibfield  {author} {\bibinfo {author} {\bibfnamefont {V.}~\bibnamefont
  {Jani\ifmmode~\check{s}\else \v{s}\fi{}}},\ }\href {\doibase
  10.1103/PhysRevB.60.11345} {\bibfield  {journal} {\bibinfo  {journal} {Phys.
  Rev. B}\ }\textbf {\bibinfo {volume} {60}},\ \bibinfo {pages} {11345}
  (\bibinfo {year} {1999})}\BibitemShut {NoStop}%
\bibitem [{\citenamefont {Janis}(2003)}]{Janis2}%
  \BibitemOpen
  \bibfield  {author} {\bibinfo {author} {\bibfnamefont {V.}~\bibnamefont
  {Janis}},\ }\href {http://stacks.iop.org/0953-8984/15/i=21/a=102} {\bibfield
  {journal} {\bibinfo  {journal} {Journal of Physics: Condensed Matter}\
  }\textbf {\bibinfo {volume} {15}},\ \bibinfo {pages} {L311} (\bibinfo {year}
  {2003})}\BibitemShut {NoStop}%
\bibitem [{\citenamefont {Phillips}\ \emph {et~al.}(2015)\citenamefont
  {Phillips}, \citenamefont {Kananenka},\ and\ \citenamefont
  {Zgid}}]{fractional}%
  \BibitemOpen
  \bibfield  {author} {\bibinfo {author} {\bibfnamefont {J.~J.}\ \bibnamefont
  {Phillips}}, \bibinfo {author} {\bibfnamefont {A.~A.}\ \bibnamefont
  {Kananenka}}, \ and\ \bibinfo {author} {\bibfnamefont {D.}~\bibnamefont
  {Zgid}},\ }\href@noop {} {\bibfield  {journal} {\bibinfo  {journal} {The
  Journal of chemical physics}\ }\textbf {\bibinfo {volume} {142}},\ \bibinfo
  {pages} {194108} (\bibinfo {year} {2015})}\BibitemShut {NoStop}%
\bibitem [{\citenamefont {Kananenka}\ \emph
  {et~al.}(2016{\natexlab{b}})\citenamefont {Kananenka}, \citenamefont
  {Phillips},\ and\ \citenamefont {Zgid}}]{legendre}%
  \BibitemOpen
  \bibfield  {author} {\bibinfo {author} {\bibfnamefont {A.~A.}\ \bibnamefont
  {Kananenka}}, \bibinfo {author} {\bibfnamefont {J.~J.}\ \bibnamefont
  {Phillips}}, \ and\ \bibinfo {author} {\bibfnamefont {D.}~\bibnamefont
  {Zgid}},\ }\href@noop {} {\bibfield  {journal} {\bibinfo  {journal} {Journal
  of chemical theory and computation}\ } (\bibinfo {year}
  {2016}{\natexlab{b}})}\BibitemShut {NoStop}%
\bibitem [{\citenamefont {Choi}\ \emph {et~al.}(2016)\citenamefont {Choi},
  \citenamefont {Kutepov}, \citenamefont {Haule}, \citenamefont {van
  Schilfgaarde},\ and\ \citenamefont {Kotliar}}]{Kotliar}%
  \BibitemOpen
  \bibfield  {author} {\bibinfo {author} {\bibfnamefont {S.}~\bibnamefont
  {Choi}}, \bibinfo {author} {\bibfnamefont {A.}~\bibnamefont {Kutepov}},
  \bibinfo {author} {\bibfnamefont {K.}~\bibnamefont {Haule}}, \bibinfo
  {author} {\bibfnamefont {M.}~\bibnamefont {van Schilfgaarde}}, \ and\
  \bibinfo {author} {\bibfnamefont {G.}~\bibnamefont {Kotliar}},\ }\href@noop
  {} {\bibfield  {journal} {\bibinfo  {journal} {NPJ Quantum Materials}\
  }\textbf {\bibinfo {volume} {1}},\ \bibinfo {pages} {16001} (\bibinfo {year}
  {2016})}\BibitemShut {NoStop}%
\bibitem [{\citenamefont {Baym}\ and\ \citenamefont
  {Kadanoff}(1961)}]{baymkadanoff}%
  \BibitemOpen
  \bibfield  {author} {\bibinfo {author} {\bibfnamefont {G.}~\bibnamefont
  {Baym}}\ and\ \bibinfo {author} {\bibfnamefont {L.~P.}\ \bibnamefont
  {Kadanoff}},\ }\href@noop {} {\bibfield  {journal} {\bibinfo  {journal}
  {Physical Review}\ }\textbf {\bibinfo {volume} {124}},\ \bibinfo {pages}
  {287} (\bibinfo {year} {1961})}\BibitemShut {NoStop}%
\bibitem [{\citenamefont {Kou}\ and\ \citenamefont
  {Hirata}(2014)}]{FCISoHirata}%
  \BibitemOpen
  \bibfield  {author} {\bibinfo {author} {\bibfnamefont {Z.}~\bibnamefont
  {Kou}}\ and\ \bibinfo {author} {\bibfnamefont {S.}~\bibnamefont {Hirata}},\
  }\href@noop {} {\bibfield  {journal} {\bibinfo  {journal} {Theoretical
  Chemistry Accounts}\ }\textbf {\bibinfo {volume} {133}},\ \bibinfo {pages}
  {1} (\bibinfo {year} {2014})}\BibitemShut {NoStop}%
\bibitem [{foo({\natexlab{b}})}]{footnote1}%
  \BibitemOpen
  \href@noop {} {} ({\natexlab{b}}),\ \bibinfo {note} {in our molecular
  calculations we have used 200 Legendre polynomials, 50,000 points on the
  imaginary frequency grid, and 5,200 imaginary-time points, corresponding to
  the full grid size before the spline interpolation technique was employed for
  the imaginary frequency grid. One can expect a reduction to ~10\% of the
  original grid size using this technique. We used the most strict criteria for
  the accuracy, corresponding to a $\delta=10^{-7}$. Additionally, the Legendre
  basis vastly reduces the cost arising from the number of $\tau$ points on the
  time grid.}\BibitemShut {Stop}%
\bibitem [{\citenamefont {Huzinaga}(1984)}]{huzinaga1985basis}%
  \BibitemOpen
  \bibfield  {author} {\bibinfo {author} {\bibfnamefont {S.}~\bibnamefont
  {Huzinaga}},\ }\href@noop {} {\emph {\bibinfo {title} {Gaussian basis sets
  for molecular calculations}}}\ (\bibinfo  {publisher} {Elsevier, Amsterdam},\
  \bibinfo {year} {1984})\BibitemShut {NoStop}%
\bibitem [{\citenamefont {Bauer}\ \emph {et~al.}(2011)\citenamefont {Bauer},
  \citenamefont {Carr}, \citenamefont {Evertz}, \citenamefont {Feiguin},
  \citenamefont {Freire}, \citenamefont {Fuchs}, \citenamefont {Gamper},
  \citenamefont {Gukelberger}, \citenamefont {Gull}, \citenamefont {Guertler},
  \citenamefont {Hehn}, \citenamefont {Igarashi}, \citenamefont {Isakov},
  \citenamefont {Koop}, \citenamefont {Ma}, \citenamefont {Mates},
  \citenamefont {Matsuo}, \citenamefont {Parcollet}, \citenamefont
  {Pawłowski}, \citenamefont {Picon}, \citenamefont {Pollet}, \citenamefont
  {Santos}, \citenamefont {Scarola}, \citenamefont {Schollwöck}, \citenamefont
  {Silva}, \citenamefont {Surer}, \citenamefont {Todo}, \citenamefont {Trebst},
  \citenamefont {Troyer}, \citenamefont {Wall}, \citenamefont {Werner},\ and\
  \citenamefont {Wessel}}]{ALPS}%
  \BibitemOpen
  \bibfield  {author} {\bibinfo {author} {\bibfnamefont {B.}~\bibnamefont
  {Bauer}}, \bibinfo {author} {\bibfnamefont {L.~D.}\ \bibnamefont {Carr}},
  \bibinfo {author} {\bibfnamefont {H.~G.}\ \bibnamefont {Evertz}}, \bibinfo
  {author} {\bibfnamefont {A.}~\bibnamefont {Feiguin}}, \bibinfo {author}
  {\bibfnamefont {J.}~\bibnamefont {Freire}}, \bibinfo {author} {\bibfnamefont
  {S.}~\bibnamefont {Fuchs}}, \bibinfo {author} {\bibfnamefont
  {L.}~\bibnamefont {Gamper}}, \bibinfo {author} {\bibfnamefont
  {J.}~\bibnamefont {Gukelberger}}, \bibinfo {author} {\bibfnamefont
  {E.}~\bibnamefont {Gull}}, \bibinfo {author} {\bibfnamefont {S.}~\bibnamefont
  {Guertler}}, \bibinfo {author} {\bibfnamefont {A.}~\bibnamefont {Hehn}},
  \bibinfo {author} {\bibfnamefont {R.}~\bibnamefont {Igarashi}}, \bibinfo
  {author} {\bibfnamefont {S.~V.}\ \bibnamefont {Isakov}}, \bibinfo {author}
  {\bibfnamefont {D.}~\bibnamefont {Koop}}, \bibinfo {author} {\bibfnamefont
  {P.~N.}\ \bibnamefont {Ma}}, \bibinfo {author} {\bibfnamefont
  {P.}~\bibnamefont {Mates}}, \bibinfo {author} {\bibfnamefont
  {H.}~\bibnamefont {Matsuo}}, \bibinfo {author} {\bibfnamefont
  {O.}~\bibnamefont {Parcollet}}, \bibinfo {author} {\bibfnamefont
  {G.}~\bibnamefont {Pawłowski}}, \bibinfo {author} {\bibfnamefont {J.~D.}\
  \bibnamefont {Picon}}, \bibinfo {author} {\bibfnamefont {L.}~\bibnamefont
  {Pollet}}, \bibinfo {author} {\bibfnamefont {E.}~\bibnamefont {Santos}},
  \bibinfo {author} {\bibfnamefont {V.~W.}\ \bibnamefont {Scarola}}, \bibinfo
  {author} {\bibfnamefont {U.}~\bibnamefont {Schollwöck}}, \bibinfo {author}
  {\bibfnamefont {C.}~\bibnamefont {Silva}}, \bibinfo {author} {\bibfnamefont
  {B.}~\bibnamefont {Surer}}, \bibinfo {author} {\bibfnamefont
  {S.}~\bibnamefont {Todo}}, \bibinfo {author} {\bibfnamefont {S.}~\bibnamefont
  {Trebst}}, \bibinfo {author} {\bibfnamefont {M.}~\bibnamefont {Troyer}},
  \bibinfo {author} {\bibfnamefont {M.~L.}\ \bibnamefont {Wall}}, \bibinfo
  {author} {\bibfnamefont {P.}~\bibnamefont {Werner}}, \ and\ \bibinfo {author}
  {\bibfnamefont {S.}~\bibnamefont {Wessel}},\ }\href
  {http://stacks.iop.org/1742-5468/2011/i=05/a=P05001} {\bibfield  {journal}
  {\bibinfo  {journal} {Journal of Statistical Mechanics: Theory and
  Experiment}\ }\textbf {\bibinfo {volume} {2011}},\ \bibinfo {pages} {P05001}
  (\bibinfo {year} {2011})}\BibitemShut {NoStop}%
\bibitem [{\citenamefont {Miro}\ \emph {et~al.}(2014)\citenamefont {Miro},
  \citenamefont {Audiffred},\ and\ \citenamefont {Heine}}]{C4CS00102H}%
  \BibitemOpen
  \bibfield  {author} {\bibinfo {author} {\bibfnamefont {P.}~\bibnamefont
  {Miro}}, \bibinfo {author} {\bibfnamefont {M.}~\bibnamefont {Audiffred}}, \
  and\ \bibinfo {author} {\bibfnamefont {T.}~\bibnamefont {Heine}},\ }\href
  {\doibase 10.1039/C4CS00102H} {\bibfield  {journal} {\bibinfo  {journal}
  {Chem. Soc. Rev.}\ }\textbf {\bibinfo {volume} {43}},\ \bibinfo {pages}
  {6537} (\bibinfo {year} {2014})}\BibitemShut {NoStop}%
\bibitem [{\citenamefont {Neese}\ and\ \citenamefont
  {Valeev}(2011)}]{doi:10.1021/ct100396y}%
  \BibitemOpen
  \bibfield  {author} {\bibinfo {author} {\bibfnamefont {F.}~\bibnamefont
  {Neese}}\ and\ \bibinfo {author} {\bibfnamefont {E.~F.}\ \bibnamefont
  {Valeev}},\ }\href {\doibase 10.1021/ct100396y} {\bibfield  {journal}
  {\bibinfo  {journal} {Journal of Chemical Theory and Computation}\ }\textbf
  {\bibinfo {volume} {7}},\ \bibinfo {pages} {33} (\bibinfo {year} {2011})},\
  \bibinfo {note} {pMID: 26606216},\ \Eprint
  {http://arxiv.org/abs/http://dx.doi.org/10.1021/ct100396y}
  {http://dx.doi.org/10.1021/ct100396y} \BibitemShut {NoStop}%
\bibitem [{\citenamefont {Lan}\ \emph {et~al.}(2015)\citenamefont {Lan},
  \citenamefont {Kananenka},\ and\ \citenamefont {Zgid}}]{LAN}%
  \BibitemOpen
  \bibfield  {author} {\bibinfo {author} {\bibfnamefont {T.~N.}\ \bibnamefont
  {Lan}}, \bibinfo {author} {\bibfnamefont {A.~A.}\ \bibnamefont {Kananenka}},
  \ and\ \bibinfo {author} {\bibfnamefont {D.}~\bibnamefont {Zgid}},\
  }\href@noop {} {\bibfield  {journal} {\bibinfo  {journal} {The Journal of
  chemical physics}\ }\textbf {\bibinfo {volume} {143}},\ \bibinfo {pages}
  {241102} (\bibinfo {year} {2015})}\BibitemShut {NoStop}%
\bibitem [{\citenamefont {Nguyen~Lan}\ \emph {et~al.}(2016)\citenamefont
  {Nguyen~Lan}, \citenamefont {Kananenka},\ and\ \citenamefont
  {Zgid}}]{nguyen2016rigorous}%
  \BibitemOpen
  \bibfield  {author} {\bibinfo {author} {\bibfnamefont {T.}~\bibnamefont
  {Nguyen~Lan}}, \bibinfo {author} {\bibfnamefont {A.~A.}\ \bibnamefont
  {Kananenka}}, \ and\ \bibinfo {author} {\bibfnamefont {D.}~\bibnamefont
  {Zgid}},\ }\href {http://dx.doi.org/10.1021/acs.jctc.6b00638} {\bibfield
  {journal} {\bibinfo  {journal} {Journal of Chemical Theory and Computation}\
  }\textbf {\bibinfo {volume} {12}},\ \bibinfo {pages} {4856} (\bibinfo {year}
  {2016})}\BibitemShut {NoStop}%
\bibitem [{\citenamefont {Rusakov}\ \emph {et~al.}(2014)\citenamefont
  {Rusakov}, \citenamefont {Phillips},\ and\ \citenamefont
  {Zgid}}]{rusakov2014local}%
  \BibitemOpen
  \bibfield  {author} {\bibinfo {author} {\bibfnamefont {A.~A.}\ \bibnamefont
  {Rusakov}}, \bibinfo {author} {\bibfnamefont {J.~J.}\ \bibnamefont
  {Phillips}}, \ and\ \bibinfo {author} {\bibfnamefont {D.}~\bibnamefont
  {Zgid}},\ }\href@noop {} {\bibfield  {journal} {\bibinfo  {journal} {The
  Journal of chemical physics}\ }\textbf {\bibinfo {volume} {141}},\ \bibinfo
  {pages} {194105} (\bibinfo {year} {2014})}\BibitemShut {NoStop}%
\end{thebibliography}
\end{document}